\documentclass[11pt]{article}
\usepackage{graphicx}
\usepackage{amsmath}
\usepackage{amssymb}
\usepackage{amsfonts}
\usepackage{caption}
\usepackage{mathtools}
\usepackage{color}
\usepackage[utf8]{inputenc}
\usepackage[english]{babel}
\usepackage[table]{xcolor}
\usepackage{appendix}
\usepackage{array}
\usepackage{cellspace}
\usepackage[left=3cm,right=3cm, top=3cm, bottom=3cm]{geometry}
\newcolumntype{L}[1]{>{\raggedright\arraybackslash} m{#1} }
\usepackage[breaklinks=true,colorlinks=true,linkcolor=blue,urlcolor=blue,citecolor=blue]{hyperref}

\begin{document}

\title{\Huge The space-time line element for static ellipsoidal objects}

  \author{\text{Ranchhaigiri Brahma$^{1\ast}$ } and \text{A.K. Sen$^{1\dagger}$  }\\
  \\ \textsl{$^1$Department of Physics, Assam University, Silchar-788011, India}\\ $^{\ast}$ Email: \href{mailto: rbrahma084@gmail.com}{\textit{ rbrahma084@gmail.com}}\\
  $^{\dagger}$ Email: \href{mailto: asokesen@yahoo.com}{\textit{ asokesen@yahoo.com}}\\
  $^{\dagger}$This author contributed equally to this work.}
\date{  }
\maketitle

\abstract{In this paper, we solved the Einstein's field equation and obtained a line element for static, ellipsoidal objects characterized by the linear eccentricity ($\eta$) instead of quadrupole parameter ($q$). This line element recovers the Schwarzschild line element when $\eta$ is zero. In addition to that it also reduces to the Schwarzschild line element, if  we neglect terms of the order of $r^{-2}$ or higher which are present within the expressions for metric elements for large distances. Furthermore, as the ellipsoidal character of the derived line element is maintained by the linear eccentricity ($\eta$), which is an easily measurable parameter, this line element could be more suitable for various analytical as well as observational studies.}

\section{Introduction}\label{sec1}
The space-time line element plays an important role in general theory of relativity (GTR). Such line element for the static and spherical objects was found by Schwarzschild \cite{Schwarzschild:1916uq} in 1916 right after the GTR has been formulated. On the other hand, Weyl \cite{Weyl:1917gp} found another form of the space-time line element for the objects which possess axial symmetry and was extended by Lewis \cite{Lewis}, Papapetrou \cite{papapetrou1953rotationssymmetrische} and others. Later that form of the line element was known as Weyl-Lewis-Papapetrou form \cite{Sloane:1978ne, Frutos1}.\\\\
For the first time in 1959 Erez and Rosen \cite{Erez} solved the Einstein's field equation for the non-spherical objects with quadrupole parameter ($q$) by using the Weyl metric. Later this solution was investigated by  Doroshkevich et al. \cite{1966JETP...22..122D}, Winicour et al. \cite{PhysRev.176.1507}, Young and Coulter \cite{PhysRev.184.1313}, Quevedo and Parkes \cite{1989GReGr..21.1047Q} etc. In addition to that few number of line elements for the non-spherical objects with quadrupole parameter ($q$) were also obtained by different authors and those have been discussed in \cite{Frutos1, stephani_kramer_maccallum_hoenselaers_herlt_2003}. One of the such solutions known as the $q-$metric \cite{QUEVEDO_2011} was obtained by performing Zipoy-Voorhees (ZV) transformation in the Schwarzschild metric \cite{PhysRevD.2.2119, zipoy}. The $q-$metric is the most compact form of the line element of this kind in terms of their mathematical structure. The other quadrupolar line elements have complicated mathematical structure which makes them disadvantageous in various analytical studies. Additionally the accurate measurement of the quadrupole moment $J_2$ and hence the quadrupole parameter $q$ (since $q=2J_2\left(R_0/M_0\right)^2$ where $R_0$ and $M_0$ are the average radius and the mass of the gravitating object respectively) is relatively difficult \cite{1989GReGr..21.1047Q, Quevedo1990MultipoleMI, Pireaux_2003} than that of the linear eccentricity $\eta$.\\\\
The objective of the current article is to derive the space-time line element outside the static ellipsoidal gravitating object without incorporating the quadrupole parameter ($q$). Such notion of finding a space-time line element for static and ellipsoidal objects was observed in \cite{2011IJPS....6.1426N}. But in that article the procedure followed by the author during the derivation of the line element seems inconsistent. In the current study we sincerely carried out the calculations to obtain the line element where we used the ellipsoidal coordinates which are suitable in present context. Consequently the ellipsoidal symmetry of the space-time is maintained by the parameter called linear eccentricity ($\eta$) instead of quadrupole parameter ($q$) used in different static, axially symmetric solutions. Such form of the line element may be simpler and easier to use in various analytical studies and astronomical measurements.\\\\
We organized the present paper as: In \textbf{section} \ref{sec2} we constructed a geometrical form of the space-time line element for the static ellipsoidal objects. Then using this form of the line element we calculated the Christoffel symbols, the Ricci tensors, the Ricci scalar and the Einstein tensors. Then in \textbf{section} \ref{sec3} we solved the Einstein's field equations. Here firstly we obtained the time-time metric coefficient of the line element denoted by $A$. The resulting expression of $A$ helps to establish a relation between the unknown metric coefficients. Consequently we solved for the other metric coefficients using this relation and at the end we found all the unknown metric coefficients. In \textbf{section} \ref{sec4} we discussed the results under various conditions. At last in \textbf{section} \ref{sec5} we concluded and summarized our current study.

\section{Construction of the line element}\label{sec2}
It is physically relevant that the space-time geometry produced by a static and ellipsoidally symmetric object is also ellipsoidally symmetric which is a kind of axially symmetric space-time. This is similar to that a centrally symmetric object produces a centrally symmetric space-time geometry (\cite{bib4landau} p. 282). Therefore in order to derive the space-time line element due to ellipsoidal gravitating objects we chose the ellipsoidal coordinate system ($u, \theta, \phi$) which is suitable in current context. 
The space-time line element in such a system of coordinates can be found by starting from the Minkowski space-time in Cartesian coordinates as below:
\begin{align}\label{eq:one}
    ds^2=c^2dt^2-dx^2-dy^2-dz^2
\end{align}
Then transforming the coordinates $(x, y, z)\longrightarrow (u, \theta, \phi)$ by the relations (\cite{bib4} pp. 39-40):
\begin{align}\label{eq:two}
    x=\sqrt{u^2+\eta^2}\sin\theta \cos\phi;\hspace{0.5cm}
    y=\sqrt{u^2+\eta^2}\sin\theta \sin\phi;\hspace{0.5cm}
    z=u\cos\theta
\end{align}
we obtained the Minkowski space-time in ellipsoidal coordinates as (\cite{KerrST} p. 15):
\begin{align}\label{eq:three}
      ds^2&=c^2dt^2-\left[\frac{u^2+\eta^2\cos^2\theta}{u^2+\eta^2}\right]du^2-\left(u^2+\eta^2\cos^2\theta\right) d\theta^2\nonumber\\
    &\hspace{4cm}-\left(u^2+\eta^2\right)\sin^2\theta d\phi^2
\end{align}
Here the parameter $\eta=\sqrt{v^2-u^2}$ is a constant known as the linear eccentricity where $v$ and $u$ represents the semi-major and the semi-minor axis of the ellipsoid respectively; and $\theta$, $\phi$ are the polar and the azimuthal angle coordinates respectively.\\\\
Therefore the line element of the curved space-time outside the ellipsoidal object characterized by the linear eccentricity ($\eta$) with mass $M$ can be written in the form as below:
\begin{align}\label{eq:four}
    ds^2&=A^2c^2dt^2-B^2\left[\frac{u^2+\eta^2\cos^2\theta}{u^2+\eta^2}\right]du^2-D_1^2\left(u^2+\eta^2\cos^2\theta\right) d\theta^2\nonumber\\
    &\hspace{4cm}-D_2^2\left(u^2+\eta^2\right)\sin^2\theta d\phi^2
\end{align}
where $A$, $B$, $D_1$ and $D_2$ are unknown metric coefficients and are functions of $(u, \theta)$ only which indicates the axially symmetric character of the metric  and possesses the boundary condition that $A, B, D_1, D_2\longrightarrow 1$ when $M\longrightarrow 0$ (i.e. if there is no mass).\\\\
It is also to be mentioned that in ellipsoidal coordinate system denoted by ($u, \theta, \phi$), the surface at $u=$ constant represents an ellipsoid with semi-minor axis $u$ (\cite{bib4} p. 41). Therefore in order to maintain the ellipsoidally symmetric character of the line element the metric coefficients $D_1$ and $D_2$ in equation \eqref{eq:four} should be equal as described in \cite{Zsigrai_2003, Racz_1992}. So, now we write as:
\begin{align}\label{eq:five}
    D_1=D_2=D
\end{align}
Then the equation \eqref{eq:four} with $D_1=D_2=D$ represents a space-time line element due to static and ellipsoidal objects \cite{Zsigrai_2003, Racz_1992} where now we defined $\eta$ as the linear eccentricity of the gravitating object with mass $M$. This definition of $\eta$ indicates that at $\eta=0$ the equation \eqref{eq:four} should recover the Schwarzschild line element.\\\\
Now we write the metric components in equation \eqref{eq:four} as follows (by putting $D_1=D_2=D$):
\begin{equation}\label{eq:six}
    \begin{aligned}
    &e^{2\nu}=A^2\\
    &e^{2\psi}=B^2\left[\frac{u^2+\eta^2\cos^2\theta}{u^2+\eta^2}\right]\\
    &e^{2\mu_1}=D^2\left(u^2+\eta^2\cos^2\theta\right)\\
    \text{and\,\,\,\,\,}&e^{2\mu_2}=D^2\left(u^2+\eta^2\right)\sin^2\theta
\end{aligned}
\end{equation}
And then we obtain from equation \eqref{eq:four} as below:
\begin{align}\label{eq:seven}
    ds^2&=e^{2\nu}c^2dt^2-e^{2\psi}du^2-e^{2\mu_1}d\theta^2-e^{2\mu_2}d\phi^2
\end{align}
Also the line element \eqref{eq:seven} can be written in the form as below: 
\begin{align}\label{eq:eight}
     ds^2=e^{\beta}\left[\chi c^2dt^2-\frac{1}{\chi}d\phi^2\right]-\frac{e^{(\psi+\mu_1)}}{\Delta^{1/2}}\left[du^2+\Delta d\theta^2\right]
\end{align}
where
\begin{align}\label{eq:nine}
    \beta=\nu+\mu_2\,\,\,\,; \hspace{1cm}\Delta=e^{2(\mu_1-\psi)}\hspace{1cm}\text{and\,\,\,\,\,} \chi=e^{\nu-\mu_2}
\end{align}
The line element \eqref{eq:eight} is known as Chandrasekhar's form (\cite{Sloane:1978ne}, \cite{bibchnadra} p. 277) of an axially symmetric space-time. The metric functions $\nu$, $\psi$, $\mu_1$ and $\mu_2$ depend only on $u$ and $\theta$ coordinates. These are the coefficients associated with each of the coordinates described by the metric tensors $g_{ij}$ in the line element, which are independent of $t$ and $\phi$ exhibiting static and axial symmetry of the space-time\footnote{Note: The English alphabet indices ($i,j, k...$) denotes to run ($0, 1, 2, 3$) that represents $ct$, $u$, $\theta$ and $\phi$ co-ordinates respectively. Metric sign convention is $(+,-,-,-)$} (i.e. $\partial g_{ij}/\partial t=0$ and $\partial g_{ij}/\partial \phi=0$) (\cite{Frutos1}, \cite{bibchnadra} p. 66). In the present context the covariant ($g_{ij}$) is diagonal and hence the elements of its contravariant is simply reciprocal to that of covariant ($g_{ij}$) i.e. $g^{ij}=\frac{1}{g_{ij}}$.
In the form of matrix the covariant ($g_{ij}$) and contravariant ($g^{ij}$) can be written from equation \eqref{eq:seven}  as below:

\begin{align}\label{eq:ten}
(g_{ij})&=\begin{pmatrix}
e^{2\nu} & 0 & 0 & 0 \\[0.3em]
0 & -e^{2\psi} & 0 & 0\\[0.3em]
0 & 0 & -e^{2\mu_1} & 0\\[0.3em]
0 & 0 & 0 &-e^{2\mu_2}
\end{pmatrix}\\
&\nonumber\label{eq:eleven}\\
(g^{ij})&=\begin{pmatrix}
e^{-2\nu} & 0 & 0 & 0 \\[0.3em]
0 & -e^{-2\psi} & 0 & 0\\[0.3em]
0 & 0 & -e^{-2\mu_1} & 0\\[0.3em]
0 & 0 & 0 &-e^{-2\mu_2}
\end{pmatrix}
\end{align}
On the other hand, in general theory of relativity the geometry of the space-time is given by the Einstein's field equation:
\begin{align}\label{eq:twelve}
    R_{ij}-\frac{1}{2}Rg_{ij}=\kappa T_{ij}
\end{align}
where $R_{ij}$ is called the Ricci tensor, $R$ is the Ricci scalar (the trace of Ricci tensor), $T_{ij}$ is the energy-momentum tensor, and $\kappa=8\pi G/c^4$ is a constant of proportionality; $G$ is the gravitational constant and $c$ is the speed of light in vacuum. The Ricci tensor is the contraction of two upper and lower middle indices of the 
Riemann curvature tensor $R^n_{ijk}=\frac{\partial \Gamma_{ij}^n}{\partial x^k}-\frac{\partial \Gamma_{ik}^n}{\partial x^j}+\Gamma_{pk}^n\Gamma_{ij}^p-\Gamma_{pj}^n\Gamma_{ik}^p$
, that is 
\begin{align}\label{eq:thirteen}
   R_{ik}&=R^n_{ink}=\frac{\partial \Gamma_{in}^n}{\partial x^k}-\frac{\partial \Gamma_{ik}^n}{\partial x^n}+\Gamma_{pk}^n\Gamma_{in}^p-\Gamma_{pn}^n\Gamma_{ik}^p
\end{align}
where the symbol $\Gamma^n_{ik}$ is known as the Christoffel symbol of second kind that is written as:
\begin{align}\label{eq:forteen}
\Gamma_{ik}^n&=\frac{1}{2}g^{np}\left[\frac{\partial g_{ip}}{\partial x^k}+\frac{\partial g_{pk}}{\partial x^i}-\frac{\partial g_{ki}}{\partial x^p}\right]
\end{align}
Therefore, firstly we obtained the Christoffel symbols of second kind by using the above metric components in equation \eqref{eq:ten} and \eqref{eq:eleven} where ( $'$ ), and ( $\Dot{}$ ) denote the derivatives w.r.t. $u$, and $\theta$ respectively. All the non-zero Christoffel symbols under the present  circumstances are (for details see in Appendix \ref{secA1}):
\begin{equation}\label{eq:fifteen}
    \begin{aligned}
    \Gamma_{00}^1&=e^{2(\nu-\psi)}\nu^{\prime},\hspace{0.7cm}\Gamma_{00}^2=e^{2(\nu-\mu_1)}\Dot{\nu}, \hspace{0.7cm}\Gamma_{11}^1=\psi^{\prime}\\
    \Gamma_{11}^2&=-e^{2(\psi-\mu_1)}\Dot{\psi},\hspace{0.7cm}\Gamma_{22}^1=-e^{2(\mu_1-\psi)}\mu_1^{\prime},\hspace{0.7cm}\Gamma_{22}^2=\Dot{\mu_1}\\
    \Gamma_{33}^1&=-e^{2(\mu_2-\psi)}\mu_2^{\prime}, \hspace{0.7cm}\Gamma_{33}^2=-e^{2(\mu_2-\mu_1)}\Dot{\mu_2}, \hspace{0.7cm}\Gamma_{10}^0=\Gamma_{01}^0=\nu^{\prime}\\
    \Gamma_{20}^0&=\Gamma_{02}^0=\Dot{\nu}, \hspace{0.7cm}\Gamma_{12}^1=\Gamma_{21}^1=\Dot{\psi}, \hspace{0.7cm}\Gamma_{12}^2=\Gamma_{21}^2=\mu_1^{\prime}\\
    \Gamma_{13}^3&=\Gamma_{31}^3=\mu_2^{\prime}, \hspace{0.7cm} \Gamma_{23}^3=\Gamma_{32}^3=\Dot{\mu_2}
\end{aligned}
\end{equation}
The zero value Christoffel symbols are:

\begin{equation}\label{eq:sixteen}
    \begin{aligned}
    \Gamma_{00}^0&=0,
    \hspace{0.7cm}\Gamma_{00}^3=0, 
    \hspace{0.7cm}\Gamma_{11}^0=0,
    \hspace{0.7cm}\Gamma_{11}^3=0
    \hspace{0.7cm}\Gamma_{22}^0=0\\
    \Gamma_{22}^3&=0,
    \hspace{0.7cm}\Gamma_{33}^0=0,
    \hspace{0.7cm}\Gamma_{12}^0=\Gamma_{21}^0=0,
    \hspace{0.7cm}\Gamma_{13}^0=\Gamma_{31}^0=0\\
    \Gamma_{33}^3&=0,
   \hspace{0.7cm} \Gamma_{23}^0=\Gamma_{32}^0=0,
    \hspace{0.7cm}\Gamma_{30}^0=\Gamma_{03}^0=0,
    \hspace{0.7cm}\Gamma_{10}^1=\Gamma_{01}^1=0\\
    \Gamma_{10}^2&=\Gamma_{01}^2=0,
    \hspace{0.7cm}\Gamma_{10}^3=\Gamma_{01}^3=0,
     \hspace{0.7cm}\Gamma_{12}^3=\Gamma_{21}^3=0\\
     \Gamma_{13}^1&=\Gamma_{31}^1=0,
    \hspace{0.7cm} \Gamma_{13}^2=\Gamma_{31}^2=0,
     \hspace{0.7cm}\Gamma_{20}^1=\Gamma_{02}^1=0\\
     \Gamma_{20}^2&=\Gamma_{02}^2=0,
     \hspace{0.7cm}\Gamma_{20}^3=\Gamma_{02}^3=0,
     \hspace{0.7cm}\Gamma_{23}^1=\Gamma_{32}^1=0\\
     \Gamma_{23}^2&=\Gamma_{32}^2=0,
     \hspace{0.7cm}\Gamma_{30}^1=\Gamma_{03}^1=0,
     \hspace{0.7cm}\Gamma_{30}^2=\Gamma_{03}^2=0\\
      \Gamma_{30}^3&=\Gamma_{03}^3=0
\end{aligned}
\end{equation}
Then using the above Christoffel symbols in equation \eqref{eq:fifteen} and \eqref{eq:sixteen} for the line element \eqref{eq:seven} we obtained the Ricci tensors as follows (for details see in Appendix \ref{subsecA2.1}):
\begin{align}
    R_{00}&=-e^{2(\nu-\psi)}\Big[\nu^{\prime\prime}+\nu^{\prime}\left(\nu^{\prime}-\psi^{\prime}+\mu_1^{\prime}+\mu_2^{\prime}\right)\Big]\nonumber\\
    &\hspace{3cm}-e^{2(\nu-\mu_1)}\Big[\Ddot{\nu}+\Dot{\nu}\left(\Dot{\nu}-\Dot{\mu_1}+\Dot{\psi}+\Dot{\mu_2}\right)\Big]\label{eq:seventeen}\\
    &\nonumber\\
    R_{11}&=\nu^{\prime\prime}+\mu_1^{\prime\prime}+\mu_2^{\prime\prime}+\left(\nu^{\prime}\right)^2+\left(\mu_1^{\prime}\right)^2+\left(\mu_2^{\prime}\right)^2-\nu^{\prime}\psi^{\prime}-\mu_1^{\prime}\psi^{\prime}-\mu_2^{\prime}\psi^{\prime}\nonumber\\
    &+e^{2(\psi-\mu_1)}\Big(\Ddot{\psi}+\Dot{\psi}\Dot{\psi}-\Dot{\psi}\Dot{\mu_1}+\Dot{\nu}\Dot{\psi}+\Dot{\mu_2}\Dot{\psi}\Big)\label{eq:eighteen}\\
    &\nonumber\\
    R_{22}&=e^{2(\mu_1-\psi)}\Big[\mu_1^{\prime\prime}-\mu_1^{\prime}\psi^{\prime}+\mu_1^{\prime}\mu_1^{\prime}+\mu_2^{\prime}\mu_1^{\prime}+\nu^{\prime}\mu_1^{\prime}\Big]-\left(\Dot{\nu}\right)\left(\Dot{\mu_1}\right)+\Ddot{\nu}\nonumber\\
     &+\Ddot{\psi}+\Ddot{\mu_2}+\left(\Dot{\mu_2}\right)^2+\left(\Dot{\nu}\right)^2+\left(\Dot{\psi}\right)^2-\left(\Dot{\mu_2}\right)\left(\Dot{\mu_1}\right)-\left(\Dot{\psi}\right)\left(\Dot{\mu_1}\right)\label{eq:nineteen}\\
     &\nonumber\\
     R_{33}&=e^{2(\mu_2-\psi)}\Big[\mu_2^{\prime\prime}+\mu_2^{\prime}\left(\mu_2^{\prime}-\psi^{\prime}+\nu^{\prime}+\mu_1^{\prime}\right)\Big]\nonumber\\
    &\hspace{3cm}+e^{2(\mu_2-\mu_1)}\Big[\Ddot{\mu_2}+\Dot{\mu_2}\left(\Dot{\mu_2}-\Dot{\mu_1}+\Dot{\nu}+\Dot{\psi}\right)\Big]\label{eq:twenty}\\
    &\nonumber\\
    \text{and\,\,\,\,}R_{12}&=\frac{\partial }{\partial \theta}\left(\nu^{\prime}+\mu_2^{\prime}\right)+\left(\Dot{\nu}
\right)\left(\nu^{\prime}\right)+\left(\Dot{\mu_2}\right)\left(\mu_2^{\prime}\right)-\left(\nu^{\prime}\right)\left(\Dot{\psi}\right)-\left(\Dot{\nu}\right)\left(\mu_1^{\prime}\right)\nonumber\\
&-\left(\mu_2^{\prime}\right)\left(\Dot{\psi}\right)-\left(\Dot{\mu_2}\right)\left(\mu_1^{\prime}\right)\label{eq:twentyone}
\end{align}
The vanishing Ricci tensors are as below:
\begin{equation}\label{eq:twentytwo}
    \begin{aligned}
    R_{01}&=R_{10}=0,\hspace{0.7cm} R_{02}=R_{20}=0, \hspace{0.7cm}R_{03}=R_{30}=0\\
     R_{13}&=R_{31}=0\hspace{0.7cm}R_{23}=R_{32}=0
\end{aligned}
\end{equation}
The Ricci scalar which is the trace of the Ricci tensors is written as below (for details see in Appendix \ref{subsecA2.2}):
\begin{align}
    R=&-2e^{-2\psi}\Big[\nu^{\prime\prime}+\nu^{\prime}\nu^{\prime}-\nu^{\prime}\psi^{\prime}+\mu_1^{\prime}\nu^{\prime}+\mu_2^{\prime}\nu^{\prime}+\mu_1^{\prime\prime}+\mu_2^{\prime\prime}\nonumber\\
    &+\left(\mu_1^{\prime}\right)^2+\left(\mu_2^{\prime}\right)^2-\mu_1^{\prime}\psi^{\prime}-\mu_2^{\prime}\psi^{\prime}+\mu_2^{\prime}\mu_1^{\prime}\Big]\nonumber\\
    &-2e^{-2\mu_1}\Big[\Ddot{\nu}+\Dot{\nu}\Dot{\nu}-\Dot{\nu}\Dot{\mu_1}+\Dot{\psi}\Dot{\nu}+\Dot{\mu_2}\Dot{\nu}+\Ddot{\psi}+\Dot{\psi}\Dot{\psi}-\Dot{\psi}\Dot{\mu_1}+\Dot{\mu_2}\Dot{\psi}\nonumber\\
     &+\Ddot{\mu_2}+\left(\Dot{\mu_2}\right)^2-\left(\Dot{\mu_2}\right)\left(\Dot{\mu_1}\right)\Big]\label{eq:twentythree}
\end{align}
In addition to that we also calculated the ($u$-$u$) and ($\theta$-$\theta$) components of Einstein tensors as follows (for details see in Appendix \ref{subsecA2.3}):
\begin{align}
G_{11}&=R_{11}-\frac{1}{2}g_{11}R\nonumber\\
     &=-\mu_1^{\prime}\nu^{\prime}-\mu_2^{\prime}\nu^{\prime}-\mu_2^{\prime}\mu_1^{\prime}-e^{2(\psi-\mu_1)}\Big[\Ddot{\nu}+\Dot{\nu}\Dot{\nu}-\Dot{\nu}\Dot{\mu_1}+\Dot{\mu_2}\Dot{\nu}\nonumber\\
     &+\Ddot{\mu_2}+\Dot{\mu_2}\Dot{\mu_2}-\Dot{\mu_2}\Dot{\mu_1}\Big]\label{eq:twentyfour}
\end{align}
and
\begin{align}
    G_{22}&=R_{22}-\frac{1}{2}g_{22}R\nonumber\\
     &=-e^{2(\mu_1-\psi)}\Big[\nu^{\prime\prime}+\nu^{\prime}\nu^{\prime}-\nu^{\prime}\psi^{\prime}+\mu_2^{\prime}\nu^{\prime}+\mu_2^{\prime\prime}+\mu_2^{\prime}\mu_2^{\prime}-\mu_2^{\prime}\psi^{\prime}\Big]\nonumber\\
     &-\Big[\Dot{\psi}\Dot{\nu}+\Dot{\mu_2}\Dot{\nu}+\Dot{\mu_2}\Dot{\psi}\Big]\label{eq:twentyfive}
\end{align}

\section{Solution of the Einstein's field equation}\label{sec3}
The objective of the current study is to obtain the line element of space-time geometry outside the ellipsoidal gravitating object. Therefore we put the energy-momentum tensor $T_{ij}=0$, and then the Einstein's field equation becomes as below:
\begin{align}\label{eq:twentysix}
    R_{ij}-\frac{1}{2}g_{ij}R=0
\end{align}
By taking trace of this equation \eqref{eq:twentysix} we obtained:
\begin{align}\label{eq:twentyseven}
    &g^{ij}R_{ij}-\frac{1}{2}g^{ij}g_{ij}R=0\Rightarrow R_i^i-\frac{1}{2}\delta_i^iR=0\nonumber\\
    \text{or\,\,\,}&R=0
\end{align}
Therefore the Einstein's field equation outside the gravitating object is simply:
\begin{align}\label{eq:twentyeight}
    R_{ij}=0
\end{align}
The equations \eqref{eq:twentyseven} and \eqref{eq:twentyeight} are the two important equations from which we can solve for the metric coefficients $A, B$ and $D$. Now re-writing equations \eqref{eq:seventeen} and \eqref{eq:twenty} we obtained:
\begin{align}
    &R_{00}=0\nonumber\\
   \text{or\,\,\,\,} &-e^{2\nu}\left[\frac{\partial}{\partial u} \left(e^{\nu-\psi+\mu_1+\mu_2}\nu^{\prime}\right)+\frac{\partial}{\partial\theta} \left(e^{\nu+\psi-\mu_1+\mu_2}\Dot{\nu}\right)\right]=0\nonumber\\
   \text{or\,\,\,\,}&\frac{\partial}{\partial u} \left(e^{\nu-\psi+\mu_1+\mu_2}\nu^{\prime}\right)+\frac{\partial}{\partial\theta} \left(e^{\nu+\psi-\mu_1+\mu_2}\Dot{\nu}\right)=0\label{eq:twentynine}
\end{align}
and
\begin{align}
    &R_{33}=0\nonumber\\
    \text{or\,\,\,\,\,}&e^{2\mu_2}\left[\frac{\partial}{\partial u} \left(e^{\nu-\psi+\mu_1+\mu_2}\mu_2^{\prime}\right)+\frac{\partial}{\partial \theta} \left(e^{\nu+\psi-\mu_1+\mu_2}\Dot{\mu_2}\right)\right]=0\nonumber\\
    \text{or\,\,\,\,\,}&\frac{\partial}{\partial u} \left(e^{\nu-\psi+\mu_1+\mu_2}\mu_2^{\prime}\right)+\frac{\partial}{\partial \theta} \left(e^{\nu+\psi-\mu_1+\mu_2}\Dot{\mu_2}\right)=0\label{eq:thirty}
\end{align}
The sum of these two equations \eqref{eq:twentynine} and \eqref{eq:thirty} ($R_{00}+R_{33}=0$) is:
\begin{align}
     &\frac{\partial}{\partial u} \left[e^{\nu-\psi+\mu_1+\mu_2}\nu^{\prime}+e^{\nu-\psi+\mu_1+\mu_2}\mu_2^{\prime}\right]\nonumber\\
     &\hspace{1cm}+\frac{\partial}{\partial\theta} \left[e^{\nu+\psi-\mu_1+\mu_2}\Dot{\nu}+e^{\nu+\psi-\mu_1+\mu_2}\Dot{\mu_2}\right]=0\nonumber\\
     \text{or\,\,\,\,\,}&\frac{\partial}{\partial u} \left[e^{\mu_1-\psi}\frac{\partial}{\partial u}\left(e^{\nu+\mu_2}\right)\right]+\frac{\partial}{\partial\theta} \left[e^{\psi-\mu_1}\frac{\partial}{\partial \theta}\left(e^{\nu+\mu_2}\right)\right]=0\label{eq:thirtyone}
\end{align}
and the difference of the equations \eqref{eq:twentynine} and \eqref{eq:thirty} ($R_{00}-R_{33}=0$) is:
\begin{align}
     &\frac{\partial}{\partial u} \left[e^{\nu-\psi+\mu_1+\mu_2}\nu^{\prime}\right]+\frac{\partial}{\partial\theta} \left[e^{\nu+\psi-\mu_1+\mu_2}\Dot{\nu}\right]\nonumber\\
     &\hspace{1cm}-\Big(\frac{\partial}{\partial u} \left[e^{\nu-\psi+\mu_1+\mu_2}\mu_2^{\prime}\right]+\frac{\partial}{\partial \theta} \left[e^{\nu+\psi-\mu_1+\mu_2}\Dot{\mu_2}\right]\Big)=0\nonumber\\
     \text{or   }&\frac{\partial}{\partial u} \left[e^{\nu-\psi+\mu_1+\mu_2}\left(\nu^{\prime}-\mu_2^{\prime}\right)\right]+\frac{\partial}{\partial\theta} \left[e^{\nu+\psi-\mu_1+\mu_2}\left(\Dot{\nu}-\Dot{\mu_2}\right)\right]=0\label{eq:thirtytwo}
\end{align}

\subsection{Solution for the metric coefficient \texorpdfstring{$A$}{A}}
In order to solve for the metric coefficient $A=e^{\nu}$ we assumed that the space-time line element \eqref{eq:four} permits an event horizon. Since there exist a time-like and a space-like Killing vector which implies that $\partial g_{ij}/\partial t=0$ and $\partial g_{ij}/\partial\phi=0$ in static and axially symmetric space-time   \cite{Frutos1}, therefore the event horizon can be defined as a two dimensional null surface that is spanned by the Killing vectors below which the space-time is regular (\cite{bibchnadra} p. 278; \cite{ podolsky2002gravitation} p. 124). So we proceeded with our calculation as in (\cite{bibchnadra} p. 278), by which the equation of the event horizon of a static and axially symmetric space-time may be written as: $\mathcal{N}(x^1, x^2)=0$ where the condition of null of this equation  (\cite{KerrST} p. 60, \cite{bibchnadra} p. 278) is:
\begin{align}\label{eq:thirtythree}
    g^{ij}\frac{\partial\mathcal{N}}{\partial x^{i}}\frac{\partial\mathcal{N}}{\partial x^{j}}=0
\end{align}
According to the our present context the equation \eqref{eq:thirtythree} can be written by using the metrics in equation \eqref{eq:seven} as follows:
\begin{align}
    &e^{-2\psi}(\mathcal{N}^{\prime})^2+e^{-2\mu_1}(\Dot{\mathcal{N}})^2=0\nonumber\\
    \text{or\,\,\,\,\,}&\Delta(\mathcal{N}^{\prime})^2+(\Dot{\mathcal{N}})^2=0\label{eq:thirtyfour}
\end{align}
Now exercising the gauge freedom we supposed that $\Delta$ is a function of $u$ only which is representing the null surface. i.e. $e^{2(\mu_1-\psi)}=\Delta(u)$.
So the equation of the null surface is given by,
\begin{align}\label{eq:thirtyfive}
    \Delta(u)=0
\end{align}
Again by the definition of the event horizon, the null surface is spanned by the Killing vectors $\frac{\partial}{\partial t}$ and $\frac{\partial}{\partial \phi}$, so the determinant of the metric of the subspace $(t, \phi)$ must vanish at null surface  (\cite{bibchnadra} p. 278; \cite{ podolsky2002gravitation} p. 124). i.e. from equation \eqref{eq:eight} we obtained as:
\begin{align}\label{eq:thirtysix}
    e^{2\beta}=0 \text{\,\,\,\,\,on the surface\,\,\,\,}\Delta(u)=0
\end{align}
Therefore with no loss of generality we may assume that (\cite{bibchnadra} p. 279):
\begin{align}\label{eq:thirtyseven}
    e^{\beta}=\Delta^{1/2} f(\theta)
\end{align}
which is a separable function of $u$ and $\theta$.\\\\
Now using this equation \eqref{eq:thirtyseven} we proceed to solved equation \eqref{eq:twentynine} as follows:
\begin{align}
    &\frac{\partial}{\partial u} \left[\Delta^{1/2}\frac{\partial}{\partial u}\left(\Delta^{1/2}f(\theta)\right)\right]+\frac{\partial}{\partial\theta} \left[\Delta^{-1/2}\frac{\partial}{\partial \theta}\left(\Delta^{1/2} f(\theta)\right)\right]=0\nonumber\\
     \text{or   }&\frac{\partial}{\partial u} \left[\Delta^{1/2}\frac{\partial}{\partial u}\left(\Delta^{1/2}\right)\right]=-\frac{1}{f(\theta)}\frac{\partial}{\partial\theta} \left[\frac{\partial}{\partial \theta}\left( f(\theta)\right)\right]=W\text{\,\,\,\,\,\,\, (a constant)}\label{eq:thirtyeight}
\end{align}
The above partial differential equation \eqref{eq:thirtyeight} of two independent variables are equal and hence separately equal to a constant $W$. Therefore we obtained two ordinary differential equations and they can be solved independently as follows:
\begin{align}
    &\frac{1}{f(\theta)}\frac{d}{d\theta} \left[\frac{d}{d \theta}\left( f(\theta)\right)\right]=-W\nonumber\\
    \text{or\,\,\,\,\,\,}& \frac{d^2}{d \theta^2}\left( f(\theta)\right)+Wf(\theta)=0\label{eq:thirtynine}
\end{align}
Here if $X=\frac{d}{d\theta}$ then we obtain an auxiliary equation of \eqref{eq:thirtynine} as:
\begin{align}
    &\left(X^2+W\right)f(\theta)=0\nonumber\\
    \text{or\,\,\,\,\,\,}&X=\pm i\sqrt{W}\label{eq:forty}
\end{align}
where the roots of this auxiliary equation is complex conjugate. So the general solution of the differential equation \eqref{eq:thirtynine} is:
\begin{align}\label{eq:fortyone}
    f(\theta)=N\sin\left(\sqrt{W}\theta+\delta\right)
\end{align}
where $N$ and $\delta$ are two unknown constants. Since $e^{\beta}$ is regular on $\Delta(u)=0$ and on the axis of symmetry $\theta=0$, so in equation \eqref{eq:fortyone} we have $\delta=0$. In addition to that the horizon $\Delta(u)=0$ is convex and hence by the condition of convexity of the horizon we otained $N=W=1$ (for the range $0\leq\theta\leq\pi$). Therefore by maintaining the regularity on the axis of symmetry and convexity of the horizon we may write as (\cite{bibchnadra} p. 279):
\begin{align}\label{eq:fortytwo}
    f(\theta)=\sin\left(\theta\right)
\end{align}
Using this result we again solved the differential equation \eqref{eq:thirtyeight} with variable $u$ as follows (by putting $W=1$):
\begin{align}
    &\frac{d}{d u} \left[\Delta^{1/2}\frac{d}{d u}\left(\Delta^{1/2}\right)\right]=1\nonumber\\
     \text{or    } &\int d \left[\frac{d}{d u}\left(\Delta\right)\right]=2\int du\nonumber\\
     \text{or    } & \left[\frac{d}{d u}\left(\Delta\right)\right]=2\left(u-m\right) \text{            \,\,\,\, where $m$ is a constant of integration.}\nonumber\\
     \text{or    } & \int d\Delta=2\int \left(u-m\right)du\nonumber\\
     \text{or    } &\Delta=u^2-2mu+\alpha^2\text{            \,\,\,\, where $\alpha^2$ is a constant of integration.}\label{eq:fortythree}
\end{align}
Therefore from equation \eqref{eq:thirtyseven} we obtained as:
\begin{align}\label{eq:fortyfour}
    e^{2\beta}&=\Delta\sin^2\theta 
\end{align}
Now using equation \eqref{eq:six} and \eqref{eq:nine} we obtained from equation \eqref{eq:fortyfour} as follows:
\begin{align}
    &e^{2\beta}=e^{2(\nu+\mu_2)}=\left(u^2-2mu+\alpha^2\right)\sin^2\theta\nonumber\\
    \text{or\,\,\,\,\,\,}&e^{2\nu}=A^2=\frac{u^2-2mu+\alpha^2}{D^2\left(u^2+\eta^2\right)}\label{eq:fortyfive}
\end{align}
and similarly from equation \eqref{eq:fortythree} we obtained as:
\begin{align}
    &\Delta=e^{2(\mu_1-\psi)}=\left(u^2-2mu+\alpha^2\right)\nonumber\\
    \text{or\,\,\,\,\,\,}&\frac{D^2}{B^2}=\frac{\left(u^2-2mu+\alpha^2\right)}{\left(u^2+\eta^2\right)}\nonumber\\
    \text{or\,\,\,\,\,\,}&D^2=\frac{B^2\left(u^2-2mu+\alpha^2\right)}{\left(u^2+\eta^2\right)}\label{eq:fortysix}
\end{align}
These two equations \eqref{eq:fortyfive} and \eqref{eq:fortysix} establish the relation between the metric coefficients as:
\begin{align}\label{eq:fortyseven}
    AD=\frac{D}{B}\text{\hspace{1.5cm} i.e.\,\,\,} A=\frac{1}{B}
\end{align}

\subsection{Solution for the metric coefficient \texorpdfstring{$D$}{D}}
The above equation \eqref{eq:fortyfive} and \eqref{eq:fortyseven} indicate that if we solved for the metric coefficient $D$ then the other coefficients such as $A$ and $B$ can be found easily. Therefore in order to obtain $D$ we proceed as follows where the required boundary conditions of the metric coefficient may be described as below:\\\\
\textit{\textbf{Boundary conditions:}}\label{BCD}
\begin{itemize}
   \item If $m=0$ then $D=1$  (condition for Minkowski space-time)
  \item If $u\longrightarrow\infty$ then $D\longrightarrow1$  (condition for asymptotic flatness)
  \item If $\eta=0$ then $D=1$ (condition for spherical symmetry)
\end{itemize}
Now from equation \eqref{eq:thirtytwo} we obtained by using \eqref{eq:nine}, \eqref{eq:fortythree} and \eqref{eq:fortyfour} as:
\begin{align}
&\frac{\partial}{\partial u} \left[e^{\nu-\psi+\mu_1+\mu_2}\left(\nu^{\prime}-\mu_2^{\prime}\right)\right]+\frac{\partial}{\partial\theta} \left[e^{\nu+\psi-\mu_1+\mu_2}\left(\Dot{\nu}-\Dot{\mu_2}\right)\right]=0\nonumber\\
     \text{or\,\,\,\,}&\frac{\partial}{\partial u} \left[e^{\beta}\Delta^{1/2}\frac{\chi^{\prime}}{\chi}\right]+\frac{\partial}{\partial\theta} \left[e^{\beta}\Delta^{-1/2}\frac{\Dot{\chi}}{\chi}\right]=0\nonumber\\
   \text{or\,\,\,\,} &\frac{\partial}{\partial u} \left[\Delta\frac{\chi^{\prime}}{\chi}\right]+\frac{1}{\sin\theta}\frac{\partial}{\partial\theta} \left[\sin\theta\frac{\Dot{\chi}}{\chi}\right]=0\label{eq:xfortyeight}
\end{align}
Then substituting the expressions from equations \eqref{eq:six}, \eqref{eq:nine} and \eqref{eq:fortyseven} in the differential equation \eqref{eq:xfortyeight} we obtained (for details see in Appendix \ref{secA3}):
\begin{align}
     &\Delta D^2\frac{\partial^2}{\partial u^2}\frac{1}{D^2}+\Delta\left(\frac{\partial}{\partial u}\frac{1}{D^2}\right)\left(\frac{\partial}{\partial u} D^2\right)+2(u-m)D^2\frac{\partial}{\partial u}\frac{1}{D^2}\nonumber\\
     &+ D^2\frac{\partial^2}{\partial \theta^2}\frac{1}{D^2}+\left(\frac{\partial}{\partial\theta} D^2\right)\left(\frac{\partial}{\partial \theta}\frac{1}{D^2}\right)+D^2\cot\theta\frac{\partial}{\partial \theta}\frac{1}{D^2}\nonumber\\
     &+2-\frac{4u(u-m)}{(u^2+\eta^2)}+\frac{4u^2\Delta}{(u^2+\eta^2)^2}-\frac{2\Delta}{(u^2+\eta^2)}=0\label{eq:fortyeight}
\end{align}
Now by executing the above \hyperref[BCD]{boundary condition} we put $\eta=0$ and $D=1$ in equation \eqref{eq:fortyeight} and obtained as:
\begin{align}
\alpha=0\label{eq:fifty}
\end{align}
The constant of integration $\alpha=0$ when $\eta=0$ indicates that, this is a parameter related to the shape of the gravitating object. Therefore we may  consider:
\begin{align}\label{eq:fiftyone}
    \alpha=\eta
\end{align}
Similarly using this \hyperref[BCD]{boundary condition} that $D=1$ when $\alpha=\eta=0$ in Schwarzschild limit we obtained from equation \eqref{eq:fortyfive} as:
\begin{align}\label{eq:fiftytwo}
    g_{00}=e^{2\nu}&=1-\frac{2m}{u}
\end{align}
This is exactly same as the time-time component of the Schwarzschild metric (\cite{bib4landau} p. 284). Hence we confirmed that the constant of integration $m$ is the half of the Schwarzschild radius ($r_g$) of the gravitating object with mass $M$. That is,
\begin{align}\label{eq:fiftythree}
    r_g=2m=\frac{2GM}{c^2}
\end{align}
where $G$ is the gravitational constant and $c$ is the speed of light in vacuum.\\\\
Therefore, putting $\alpha=\eta$ the equation \eqref{eq:fortyeight} can be written as:
\begin{align}
    &\Delta D^2\frac{\partial^2}{\partial u^2}\frac{1}{D^2}+\Delta\left(\frac{\partial}{\partial u}\frac{1}{D^2}\right)\left(\frac{\partial}{\partial u} D^2\right)+2(u-m)D^2\frac{\partial}{\partial u}\frac{1}{D^2}\nonumber\\
     &+ D^2\frac{\partial^2}{\partial \theta^2}\frac{1}{D^2}+\left(\frac{\partial}{\partial\theta} D^2\right)\left(\frac{\partial}{\partial \theta}\frac{1}{D^2}\right)+D^2\cot\theta\frac{\partial}{\partial \theta}\frac{1}{D^2}\nonumber\\
     &+\frac{8m\eta^2u}{(u^2+\eta^2)^2}=0\label{eq:fiftyfour}
\end{align}
In equation \eqref{eq:fiftyfour} it is observed that the partial differential equation does not contain any cross terms of variables $u$ and $\theta$ and hence this can be solved by separation of variables as calculating in (\cite{bibchnadra} pp. 344-345) and  \cite{PhysRev.174.1559}. So we considered,
\begin{align}
    D=e^{\eta\lambda(u, \theta)}\text{\,\,;\hspace{1cm}and,\,\,}\lambda(u,\theta)=P(u)+Q(\theta)\label{eq:fiftyfive}
\end{align}
where $P(u)$ and $Q(\theta)$ are respectively functions of $u$ and $\theta$ only. Substituting from equation \eqref{eq:fiftyfive} into \eqref{eq:fiftyfour} and simplifying we obtained as below :
\begin{align}
    &4\eta^2\Delta\left(\frac{\partial\lambda}{\partial u}\right)^2-2\eta \Delta\frac{\partial^2\lambda}{\partial u^2}-4\eta^2\Delta\left( \frac{\partial\lambda}{\partial u}\right)^2-4(u-m)\eta\left( \frac{\partial\lambda}{\partial u}\right)+\frac{8m\eta^2u}{(u^2+\eta^2)^2}\nonumber\\
    &+ 4\eta^2\left(\frac{\partial\lambda}{\partial \theta}\right)^2-2\eta \frac{\partial^2\lambda}{\partial \theta^2}-4\eta^2\left( \frac{\partial\lambda}{\partial\theta} \right)^2-2\eta\cot\theta\left( \frac{\partial\lambda}{\partial \theta}\right)=0\nonumber\\
    \text{or\,\,\,\,}  &2\eta \Delta\frac{\partial^2P}{\partial u^2}+4(u-m)\eta\left( \frac{\partial P}{\partial u}\right)-\frac{8m\eta^2u}{(u^2+\eta^2)^2}\nonumber\\
    &\hspace{3cm}=-2\eta \frac{\partial^2Q}{\partial \theta^2}-2\eta\cot\theta\left( \frac{\partial Q}{\partial \theta}\right)\label{eq:xfiftyseven}
\end{align}
Since the left hand-side of the above equation \eqref{eq:xfiftyseven} depend only on variable $u$, while the right hand-side only depend on the variable $\theta$ and hence they are separately equal to a constant $K$ which can be written as below:
\begin{align}
     &2\eta \Delta\frac{\partial^2P}{\partial u^2}+4(u-m)\eta\left( \frac{\partial P}{\partial u}\right)-\frac{8m\eta^2u}{(u^2+\eta^2)^2}\nonumber\\
    &\hspace{3cm}=-2\eta \frac{\partial^2Q}{\partial \theta^2}-2\eta\cot\theta\left( \frac{\partial Q}{\partial \theta}\right)=K\label{eq:fiftysix}
\end{align}
Therefore we obtained two ordinary differential equations from equation \eqref{eq:fiftysix} as below:
\begin{align}
    &2\eta \Delta\frac{d^2P}{d u^2}+4(u-m)\eta\left( \frac{d P}{d u}\right)=K+\frac{8m\eta^2u}{(u^2+\eta^2)^2}\label{eq:fiftyeight}\\
    \text{and\,\,\,}&2\eta \frac{\partial^2Q}{\partial \theta^2}+2\eta\cot\theta\left( \frac{\partial Q}{\partial \theta}\right)=-K\label{eq:fiftynine}
\end{align}
The solution of equation \eqref{eq:fiftyeight} is:
\begin{align}
    P(u)=&\frac{1}{2\eta}\Bigg[\frac{\arctan\left(\frac{u-m}{(\eta^2-m^2)^{1/2}}\right)\left(Km-2m+2\eta a_1\right)}{(\eta^2-m^2)^{1/2}}-\ln\left(u^2+n^2\right)\nonumber\\
    &+\frac{1}{2}(2+K)\ln\left(u^2-2mu+\eta^2\right)\Bigg]+a_2]\label{eq:sixty}
\end{align}
where $a_1$ and $a_2$ are constants of integration.\\\\
Similarly, the solution of equation \eqref{eq:fiftynine} is:
\begin{align}
    Q(\theta)=\frac{K}{2\eta}\ln(\sin\theta)-a_3\Big[\ln\left(\cos(\theta/2)\right)-\ln\left(\sin(\theta/2)\right)\Big]+a_4\label{eq:sixtyone}
\end{align}
where $a_3$ and $a_4$ are constants of integration.\\\\
Therefore using \eqref{eq:sixty} and \eqref{eq:sixtyone} we obtained from \eqref{eq:fiftyfive} as:
\begin{align*}
    \eta\lambda(u,\theta)=&\eta\left(P(u)+Q(\theta)\right)\nonumber\\
    =&\frac{1}{2}\Bigg[\frac{\arctan\left(\frac{u-m}{(\eta^2-m^2)^{1/2}}\right)\left(Km-2m+2\eta a_1\right)}{(\eta^2-m^2)^{1/2}}-\ln\left(u^2+n^2\right)\nonumber\\
    &+\frac{1}{2}(2+K)\ln\left(u^2-2mu+\eta^2\right)\Bigg]+\eta a_2+\frac{K}{2}\ln(\sin\theta)\nonumber\\
    &-\eta a_3\Big[\ln\left(\cos(\theta/2)\right)-\ln\left(\sin(\theta/2)\right)\Big]+\eta a_4\nonumber\\
    \end{align*}
    \begin{align}
    =&\frac{1}{2}\Bigg[\frac{\arctan\left\{\frac{u\left(1-\frac{m}{u}\right)}{(\eta^2-m^2)^{1/2}}\right\}\left(Km-2m+2\eta a_1\right)}{(\eta^2-m^2)^{1/2}}-\ln\left(u^2\right)-\ln\left(1+\frac{\eta^2}{u^2}\right)\nonumber\\
    &+\frac{1}{2}(2+K)\left\{\ln(u^2)+\ln\left(1-\frac{2m}{u}+\frac{\eta^2}{u^2}\right)\right\}\Bigg]+\eta a_2+\frac{K}{2}\ln(\sin\theta)\nonumber\\
    &-\eta a_3\Big[\ln\left(\cos(\theta/2)\right)-\ln\left(\sin(\theta/2)\right)\Big]+\eta a_4\nonumber\\
    =&\frac{1}{2}\Bigg[\frac{\arctan\left\{\frac{u\left(1-\frac{m}{u}\right)}{(\eta^2-m^2)^{1/2}}\right\}\left(Km-2m+2\eta a_1\right)}{(\eta^2-m^2)^{1/2}}+\frac{K}{2}\ln\left(u^2\right)-\ln\left(1+\frac{\eta^2}{u^2}\right)\nonumber\\
    &+\frac{1}{2}(2+K)\left\{\ln\left(1-\frac{2m}{u}+\frac{\eta^2}{u^2}\right)\right\}\Bigg]+\eta a_2+\frac{K}{2}\ln(\sin\theta)\nonumber\\
    &-\eta a_3\Big[\ln\left(\cos(\theta/2)\right)-\ln\left(\sin(\theta/2)\right)\Big]+\eta a_4\label{eq:sixtytwo}
\end{align}
The constant parameters in equation \eqref{eq:sixtytwo} are obtained by applying the \hyperref[BCD]{boundary conditions} for $D$ as follows:\\\\ 
\textbf{(a) Condition for Minkowski space-time:} If $m=0$ then $D=1$ i.e. $\eta\lambda=0$. Therefore from equation \eqref{eq:sixtytwo} we obtained as:
\begin{align}
    \eta a_2+\eta a_4=&-\frac{1}{2}\Bigg[\arctan\left\{\frac{u}{\eta}\right\}\left(2 a_1\right)+\frac{K}{2}\ln\left(u^2\right)+\frac{K}{2}\left\{\ln\left(1+\frac{\eta^2}{u^2}\right)\right\}\Bigg]\nonumber\\
    &-\frac{K}{2}\ln(\sin\theta)+\eta a_3\Big[\ln\left(\cos(\theta/2)\right)-\ln\left(\sin(\theta/2)\right)\Big]\label{eq:sixtythree}
\end{align}
Substituting this into equation \eqref{eq:sixtytwo} we obtained:
\begin{align*}
    \eta\lambda=&\frac{1}{2}\Bigg[\frac{\arctan\left\{\frac{u\left(1-\frac{m}{u}\right)}{(\eta^2-m^2)^{1/2}}\right\}\left(Km-2m+2\eta a_1\right)}{(\eta^2-m^2)^{1/2}}+\frac{K}{2}\ln\left(u^2\right)\nonumber\\
    &-\ln\left(1+\frac{\eta^2}{u^2}\right)+\frac{1}{2}(2+K)\left\{\ln\left(1-\frac{2m}{u}+\frac{\eta^2}{u^2}\right)\right\}\Bigg]\nonumber\\
    &-\frac{K}{2}\ln(\sin\theta)+\eta a_3\left[\ln\cos(\theta/2)-\ln\sin(\theta/2)\right]\nonumber\\
    &+\frac{K}{2}\ln(\sin\theta)-\eta a_3\left[\ln\cos(\theta/2)-\ln\sin(\theta/2)\right]\nonumber\\
    &-\frac{1}{2}\Bigg[\arctan\left\{\frac{u}{(\eta^2)^{1/2}}\right\}\left(2 a_1\right)+\frac{K}{2}\ln\left(u^2\right)\nonumber\\
    &+\frac{K}{2}\left\{\ln\left(1+\frac{\eta^2}{u^2}\right)\right\}\Bigg]
    \end{align*}
    \begin{align}
    \text{or\,\,\,\,}\eta\lambda=&\frac{1}{2}\Bigg[\frac{\arctan\left\{\frac{u\left(1-\frac{m}{u}\right)}{(\eta^2-m^2)^{1/2}}\right\}\left(Km-2m+2\eta a_1\right)}{(\eta^2-m^2)^{1/2}}\nonumber\\
    &-\arctan\left\{\frac{u}{(\eta^2)^{1/2}}\right\}\left(2 a_1\right)+\frac{1}{2}(2+K)\left\{\ln\left(1-\frac{2m}{u}+\frac{\eta^2}{u^2}\right)\right\}\nonumber\\
    &-\frac{1}{2}(2+K)\left\{\ln\left(1+\frac{\eta^2}{u^2}\right)\right\}\Bigg]\label{eq:sixtyfour}
\end{align}
\textbf{(b) Condition for asymptotic flatness:}
If $u\longrightarrow\infty$ then $D=1$ i.e. $\eta\lambda=0$. So the equation \eqref{eq:sixtyfour} become:
\begin{align}
    &\frac{1}{2}\Bigg[\frac{\arctan\left\{\frac{u}{(\eta^2-m^2)^{1/2}}\right\}\left(Km-2m+2\eta a_1\right)}{(\eta^2-m^2)^{1/2}}-\arctan\left\{\frac{u}{(\eta^2)^{1/2}}\right\}\left(2 a_1\right)\Bigg]=0\nonumber\\
    \text{or\,\,\,\,}&\arctan\left\{\frac{u}{(\eta^2)^{1/2}}\right\}\left(2 a_1\right)=\Bigg[\frac{\arctan\left\{\frac{u}{(\eta^2-m^2)^{1/2}}\right\}\left(Km-2m+2\eta a_1\right)}{(\eta^2-m^2)^{1/2}}\Bigg]\label{eq:sixtyfive}
\end{align}
Again by substituting \eqref{eq:sixtyfive} into equation \eqref{eq:sixtyfour} and simplifying we found as below:
\begin{align}
    \eta\lambda=&\frac{1}{2}\Bigg[\frac{\left(Km-2m+2\eta a_1\right)}{(\eta^2-m^2)^{1/2}}\Bigg\{\arctan\left(\frac{u\left(1-\frac{m}{u}\right)}{(\eta^2-m^2)^{1/2}}\right)\nonumber\\
    &-\arctan\left(\frac{u}{(\eta^2-m^2)^{1/2}}\right)\Bigg\}+\frac{1}{2}(2+K)\left\{\ln\left(1-\frac{2m}{u}+\frac{\eta^2}{u^2}\right)\right\}\nonumber\\
    &-\frac{1}{2}(2+K)\left\{\ln\left(1+\frac{\eta^2}{u^2}\right)\right\}\Bigg]\nonumber\\
    =&\frac{1}{2}\Bigg[\frac{\left(Km-2m+2\eta a_1\right)}{(\eta^2-m^2)^{1/2}}\left\{\arctan\left(\frac{-m}{(\eta^2-m^2)^{1/2}\left(1+\frac{u(u-m)}{(\eta^2-m^2)}\right)}\right)\right\}\nonumber\\
    &+\frac{1}{2}(2+K)\left\{\ln\left(1-\frac{2m}{u}+\frac{\eta^2}{u^2}\right)\right\}-\frac{1}{2}(2+K)\left\{\ln\left(1+\frac{\eta^2}{u^2}\right)\right\}\Bigg]\label{eq:sixtysix}
\end{align}
\textbf{(c) Condition for spherical symmetry:}
If $\eta=0$ then $D=1$ i.e. $\eta\lambda=0$. So the equation \eqref{eq:sixtysix} become:
\begin{align}
    0=&\frac{1}{2}\Bigg[\frac{\left(Km-2m+2\eta a_1\right)}{(\eta^2-m^2)^{1/2}}\left\{\arctan\left(\frac{-m}{(\eta^2-m^2)^{1/2}\left(1+\frac{u(u-m)}{(\eta^2-m^2)}\right)}\right)\right\}\nonumber\\
    &+\frac{1}{2}(2+K)\left\{\ln\left(1-\frac{2m}{u}+\frac{\eta^2}{u^2}\right)\right\}-\frac{1}{2}(2+K)\left\{\ln\left(1+\frac{\eta^2}{u^2}\right)\right\}\Bigg]\label{eq:sixtyseven}
\end{align}
Here the constant of integration $a_1$ can not be a function of $u$ and the constant $K$ can not be a function of $(u, \theta)$. Therefore the terms in the above equation \eqref{eq:sixtyseven} must be separately zero. This can be achieved by the choice of $K=-2$ and $a_1=2m/\eta$.\\\\
Therefore by putting these value $K=-2$ and $a_1=2m/\eta$ into equation \eqref{eq:sixtysix} we obtained:
\begin{align}
&\eta\lambda=0\label{eq:sixtyeight}
\end{align}
and hence the metric coefficient $D$ is not a function, instead a constant which is obtained from equation \eqref{eq:fiftyfive}, i.e.
\begin{align}
    D=e^{\eta\lambda}=1\label{eq:sixtynine}
\end{align}

\section{Discussion of results}\label{sec4}
We obtained the value of the metric coefficient $D$  present in equation \eqref{eq:sixtynine} and consequently the other metric coefficients from equation \eqref{eq:five} and \eqref{eq:fortyseven} are as below:
\begin{align}
    &A^2=\frac{1}{B^2}=\frac{u^2-2mu+\eta^2}{u^2+\eta^2}\label{eq:seventy}\\
    \text{and\,\,\,\,\,\,}&D_1=D_2=D=1\label{eq:seventyone}
\end{align}
Using these metric coefficients we obtained the space-time line element for the oblate shaped the gravitating object. The line element can be written from equations \eqref{eq:four}, \eqref{eq:seventy} and \eqref{eq:seventyone} as below:
\begin{align}
    ds^2&=\frac{u^2-2mu+\eta^2}{u^2+\eta^2}c^2dt^2-\left[\frac{u^2+\eta^2\cos^2\theta}{u^2-2mu+\eta^2}\right]du^2-\left(u^2+\eta^2\cos^2\theta\right) d\theta^2\nonumber\\
    &\hspace{4.5cm}-\left(u^2+\eta^2\right)\sin^2\theta d\phi^2\label{eq:seventytwo}
\end{align}
We analyzed some important features of this line element below:
\begin{itemize}
    \item \textbf{Line element in terms of radial distance :} The derived line element can be written in terms of a radial distance ($r$) that is related to $u$ by the relation (from \eqref{eq:two}):
\begin{align}\label{eq:seventythree}
    r^2=u^2+\eta^2\sin^2\theta
\end{align}
such that
\begin{align}\label{eq:seventyfour}
    du=\frac{r\,\,dr}{\sqrt{r^2-\eta^2\sin^2\theta}}+\frac{\eta^2\sin\theta\cos\theta d\theta}{\sqrt{r^2-\eta^2\sin^2\theta}}
\end{align}
Now using equation \eqref{eq:seventythree} and \eqref{eq:seventyfour} we obtained the line element in terms of ($r$) from equation \eqref{eq:seventytwo} as follows:

\begin{align}
    ds^2&=\left[1-\frac{2m(r^2-\eta^2\sin^2\theta)^{1/2}}{\left(r^2+\eta^2\cos^2\theta\right)}\right]c^2dt^2\nonumber\\
    &-\left[\frac{r^2+\eta^2\left(\cos^2\theta-\sin^2\theta\right)}{r^2+\eta^2\cos^2\theta-2m\left(r^2-\eta^2\sin^2\theta\right)^{1/2}}\right]\frac{r^2\,\,dr^2}{r^2-\eta^2\sin^2\theta}\nonumber\\
    &-\left[1+\left\{\frac{\frac{\eta^4\sin^2\theta\cos^2\theta }{r^2-\eta^2\sin^2\theta}}{r^2+\eta^2\cos^2\theta-2m\left(r^2-\eta^2\sin^2\theta\right)^{1/2}}\right\}\right]\nonumber\\
    &\times\left(r^2+\eta^2\left(\cos^2\theta-\sin^2\theta\right)\right) d\theta^2\nonumber\\
    &-\left(r^2+\eta^2\cos^2\theta\right)\sin^2\theta d\phi^2\label{eq:seventyfive}
\end{align}

 \item \textbf{Schwarzschild limit:} 
If we put $\eta=0$ in equation \eqref{eq:seventyfive} the line element reduces to the Schwarzschild line element:
\begin{align}\label{eq:seventysix}
    ds^2&=\left(1-\frac{2m}{r}\right)c^2 dt^2-\left(1-\frac{2m}{r}\right)^{-1}dr^2-r^2 d\theta^2-r^2\sin^2\theta d\phi^2
\end{align}
Therefore it can be inferred that the parameter $\eta$ maintains the ellipsoidal character of the line element. \\
 \item \textbf{Asymptotic behaviour:} Now we rewrite the equation \eqref{eq:seventyfive} in the form as below:
\begin{align}
    ds^2&=\left[1-\frac{2m\left(1-\frac{\eta^2}{r^2}\sin^2\theta\right)^{1/2}}{r\left(1+\frac{\eta^2}{r^2}\cos^2\theta\right)}\right]c^2dt^2\nonumber\\
    &-\left[\frac{1+\frac{\eta^2}{r^2}\left(\cos^2\theta-\sin^2\theta\right)}{1+\frac{\eta^2}{r^2}\cos^2\theta-\frac{2m}{r}\left(1-\frac{\eta^2}{r^2}\sin^2\theta\right)^{1/2}}\right]\frac{dr^2}{1-\frac{\eta^2}{r^2}\sin^2\theta}\nonumber\\
    &-\left[1+\left\{\frac{\frac{\eta^4\sin^2\theta\cos^2\theta }{r^4\left(1-\frac{\eta^2}{r^2}\sin^2\theta\right)}}{\left(1+\frac{\eta^2}{r^2}\cos^2\theta-\frac{2m}{r}\left(1-\frac{\eta^2}{r^2}\sin^2\theta\right)^{1/2}\right)}\right\}\right]\nonumber\\
    &\times r^2\left(1+\frac{\eta^2}{r^2}\left(\cos^2\theta-\sin^2\theta\right)\right) d\theta^2\nonumber\\
    &-r^2\left(1+\frac{\eta^2}{r^2}\cos^2\theta\right)\sin^2\theta d\phi^2\label{eq:seventyseven}
\end{align}
Here if the value of $r$ is sufficiently large such that the terms with $r^2$ or higher powers of $r$ in the denominator can be neglected (or put zero), then we obtained from equation \eqref{eq:seventyseven} as:
\begin{align}
    ds^2=&\left[1-\frac{2m}{r}\right]c^2 dt^2-\left[\frac{1}{1-\frac{2m}{r}}\right]\,\,dr^2- r^2d\theta^2-r^2\sin^2\theta d\phi^2\label{eq:seventyeight}
\end{align}
This is exactly same as the line element of the spherically symmetric space-time. From this result it is clear that at sufficiently large distance the space-time geometry due to an ellipsoidal object behaves like that due to a spherical symmetric object. Further, at $r\longrightarrow\infty$ the line element \eqref{eq:seventyseven} reduces to the Minkowski metric which indicates that the space-time at large distance from the source is flat. So the derived line element satisfies the condition of asymptotic flatness. \\\\
In addition to that if $m=0$ then the line element \eqref{eq:seventytwo} reduces to the Minkowski space-time element in ellipsoidal coordinates as shown in equation \eqref{eq:three}. Hence it can be inferred that the parameter $m$ contains the total mass of the gravitating object.\\

\item \textbf{Physical interpretation of $\eta$:} In \textbf{section} \ref{sec2} we defined $\eta$ as the linear eccentricity of the gravitating object which ultimately means that if the shape of the object is spherical then this parameter is zero. This definition of $\eta$ can be further interpreted in terms of mass of the gravitating object as follows:\\\\
It is obvious that the total volume ($\mathbf{V}$) and the total mass ($\mathbf{M}$) of an ellipsoidal oblate object with semi-major axis ($a$) and semi-minor axis ($b$) and having mass density $\rho$ are:
\begin{align}\label{eq:seventynine}
    &\mathbf{V}=\frac{4}{3}\pi a^2b \hspace{1cm}\text{and\,\,\,\,\,\,\,\,}\mathbf{M}=\rho \mathbf{V}
\end{align}
Therefore if $\eta$ is the linear eccentricity of the ellipsoidal shape object then its semi-major axis is $a=\sqrt{b^2+\eta^2}$ and the total mass is:
\begin{align}\label{eq:eighty}
    \mathbf{M}&=\frac{4}{3}\pi\rho b\left(b^2+\eta^2\right)=\mathbf{M}_b\left(1+\frac{\eta^2}{b^2}\right)
\end{align}
where $\mathbf{M}_b=\frac{4}{3}\pi\rho b^3$ is the mass of the sphere with radius $b$. Further simplifying equation \eqref{eq:eighty} we obtained:
\begin{align}\label{eq:eightyone}
    \eta^2=\frac{b^2\left(\mathbf{M}-\mathbf{M}_b\right)}{\mathbf{M}_b}=\frac{b^2\Omega}{\mathbf{M}_b}
\end{align}
in which $\Omega=\mathbf{M}-\mathbf{M}_b$ indicates the mass that is spanned outside the sphere of radius $b$. Therefore the parameter $\eta$ is also related to the mass of the gravitating object.\\
\item \textbf{Singularity and Horizons:} The singularity of the space-time is the point at which the line element becomes undefined. In the present context the space-time horizon can be studied from equation \eqref{eq:thirtyfive} i.e. $\Delta=0$ which can be written as below in terms of $r$:
\begin{align}\label{eq:eightytwo}
    r^2+\eta^2\cos^2\theta-2m\left(r^2-\eta^2\sin^2\theta\right)^{1/2}=0
\end{align}
The roots of this equation are as presented is a single equation below:
\begin{align}\label{eq:eightythree}
    r=&\pm\sqrt{2m^2-\eta^2\cos^2\theta\pm2m\sqrt{m^2-\eta^2}}
\end{align}
From equation \eqref{eq:eightythree} it can be inferred that the space-time geometry due to an ellipsoidal object possesses two event horizons where both of them are non-spherical. In addition to that the space-time exhibits naked singularity under some conditions, e.g. at $m^2<\eta^2$. However the gravitational singularity can be identified from the study of curvature invariants such as Kretschmann scalar which is expressed as $ K=R_{ijkl}R^{ijkl}$.\\

\item \textbf{Comparison with some existing solutions:} 
In 2011 Nikouravan \cite{2011IJPS....6.1426N} claimed a solution to the space-time produced by the ellipsoidal objects which is written as follows:
\begin{align}\label{eq:eightyfour}
    ds^2=&\left(1-\frac{2m}{u}\right)c^2dt^2-\frac{1}{\left(1-\frac{2m}{u}\right)}\left[\frac{u^2+\eta^2\cos^2 \theta}{u^2+\eta^2}\right]du^2\nonumber\\
    &-\left(u^2+\eta^2\cos^2\theta\right)d\theta^2-\left(u^2+\eta^2\right)\sin^2\theta d\phi^2
\end{align}
This line element \eqref{eq:eightyfour} does not coincides with the one we obtained in  the present paper (equation \eqref{eq:seventytwo}). But it is to note that equation \eqref{eq:eightyfour} also reduces to Schwarzschild line element at sufficiently large value of $u$ and possesses asymptotically flat behaviour. The mathematical structure of this equation seems relevant, however in our opinion the author didn't state properly the physics behind the derivation of the line element. Therefore it remained unclear whether the line element presented in \cite{2011IJPS....6.1426N} really signifies the space-time produced by the ellipsoidal objects or not.\\\\
On the other hand, the $q-$metric which is the simplest form of the line elements based on the quadrupole parameter is written as below \cite{QUEVEDO_2011}: 
\begin{align}
    ds^2&=\left(1-\frac{2m}{r}\right)^{1+q}c^2dt^2-\left(1-\frac{2m}{r}\right)^{-q}\Bigg[\left(1+\frac{m^2\sin^2\theta}{r^2-2mr}\right)^{-q(2+q)}\nonumber\\
    &\hspace{4cm}\left(\frac{dr^2}{1-2m/r}+r^2d\theta^2\right)+r^2\sin^2\theta d\phi^2\Bigg]\label{eq:eightyfive}
\end{align}
Here it is obvious that the line element in equation \eqref{eq:seventytwo} will be different from \eqref{eq:eightyfive} as the former is based on linear eccentricity ($\eta$) while the later is based on quadrupole parameter ($q$). However some transformation equations may exist which can connect these two solutions. Apart from that, the $q-$metric  possesses naked singularities \cite{QUEVEDO_2011, Toktarbay_2022} whereas the equation \eqref{eq:seventytwo} exhibits naked singularities only under certain conditions. Similar to that the other quadrupolar line elements also exhibit naked singularities only at certain specific conditions as discussed in \cite{Frutos1}.

\end{itemize}
From the above discussion we conclude that the line element \eqref{eq:seventytwo} describes the space-time geometry produced by the ellipsoidal objects in the most effective manner.

\section{Conclusion}\label{sec5}
In this paper we solved the Einstein's field equation in order to obtain the line element outside the static, ellipsoidal object. Consequently we obtained a line element for the said space-time geometry which is characterized by the parameter $(\eta)$ called linear eccentricity. It is observed that this line element possesses asymptotically flat behaviour which means that at spatial infinity the line element reduces to the Minkowski space-time (i.e. the gravitational field at large distance is negligible). In addition to that the derived line element takes the form of the Schwarzschild line element when  we neglect terms in the order of $r^{-2}$  or higher in the expression for metric  elements at large distances.  i.e. the gravitational fields behaves like that of spherical symmetry in that particular distance. Further the space-time characterized by the parameter $\eta$ also exhibits naked singularities under some conditions (e.g. $m^2<\eta^2$) like as observed in quadrupolar line elements \cite{Frutos1, QUEVEDO_2011}. Since, the condition for the naked singularity for our line element is  $m<\eta$, from that it can be stated that all the celestial objects whose Schwarzschild radius is less than twice of their linear eccentricity possesses a naked singularity.\\\\
The line element derived in the present work consists of physically relevant characters and therefore we conclude that this represents most effectively the gravitational field produced by the static ellipsoidal objects with linear eccentricity $\eta$. Additionally as the derived line element contains easily measurable parameters, so it will be suitable in various theoretical studies as well as experimental measurements. 
\vspace{0.5cm}\\
\textbf{Acknowledgments:}
The author Ranchhaigiri Brahma acknowledged the Ministry of Tribal Affairs, Govt. of India for supporting to carryout research work via NFST fellowship (201920-NFST-ASS-00678).\\\\
\section*{Declaration}
\textbf{Data Availability:} Since this is a theoretical work data avaiability is not applicable and no data has been used or analysed throughout the work.
\vspace{1cm}

\begin{appendices}

\section{Christoffel symbols}\label{secA1}
The Christoffel symbols of second kind obtained by using the metric components in equation \eqref{eq:ten} and \eqref{eq:eleven} are as below (where ( $'$ ), and ( $\Dot{}$ ) denote the derivatives w.r.t. $u$, and $\theta$ respectively).
\begin{align*}
    \Gamma_{00}^0&=\frac{1}{2}g^{0p}\left[\frac{\partial g_{0p}}{\partial x^0}+\frac{\partial g_{p0}}{\partial x^0}-\frac{\partial g_{00}}{\partial x^p}\right]=\frac{1}{2}g^{00}\left[\frac{\partial g_{00}}{\partial x^0}+\frac{\partial g_{00}}{\partial x^0}-\frac{\partial g_{00}}{\partial x^0}\right]=0\\
    \Gamma_{00}^1&=\frac{1}{2}g^{1p}\left[\frac{\partial g_{0p}}{\partial x^0}+\frac{\partial g_{p0}}{\partial x^0}-\frac{\partial g_{00}}{\partial x^p}\right]=\frac{1}{2}g^{11}\left[\frac{\partial g_{01}}{\partial x^0}+\frac{\partial g_{10}}{\partial x^0}-\frac{\partial g_{00}}{\partial x^1}\right]\\
    &=-\frac{1}{2}g^{11}\frac{\partial g_{00}}{\partial x^1}=-\frac{1}{2}\left(-e^{-2\psi}\right)\frac{\partial}{\partial u}\left(e^{2\nu}\right)\\
    &=e^{2(\nu-\psi)}\nu^{\prime}\\
    \Gamma_{00}^2&=\frac{1}{2}g^{2p}\left[\frac{\partial g_{0p}}{\partial x^0}+\frac{\partial g_{p0}}{\partial x^0}-\frac{\partial g_{00}}{\partial x^p}\right]=\frac{1}{2}g^{22}\left[\frac{\partial g_{02}}{\partial x^0}+\frac{\partial g_{20}}{\partial x^0}-\frac{\partial g_{00}}{\partial x^2}\right]\\
    &=-\frac{1}{2}g^{22}\frac{\partial g_{00}}{\partial x^2}=-\frac{1}{2}\left(-e^{-2\mu_1}\right)\frac{\partial }{\partial \theta}\left(e^{2\nu}\right)\\
    &=e^{2(\nu-\mu_1)}\Dot{\nu}\\
    \Gamma_{00}^3&=\frac{1}{2}g^{3p}\left[\frac{\partial g_{0p}}{\partial x^0}+\frac{\partial g_{p0}}{\partial x^0}-\frac{\partial g_{00}}{\partial x^p}\right]=\frac{1}{2}g^{33}\left[\frac{\partial g_{03}}{\partial x^0}+\frac{\partial g_{30}}{\partial x^0}-\frac{\partial g_{00}}{\partial x^3}\right]=0\\
    \Gamma_{11}^0&=\frac{1}{2}g^{0p}\left[\frac{\partial g_{1p}}{\partial x^1}+\frac{\partial g_{p1}}{\partial x^1}-\frac{\partial g_{11}}{\partial x^p}\right]=\frac{1}{2}g^{00}\left[\frac{\partial g_{10}}{\partial x^1}+\frac{\partial g_{01}}{\partial x^1}-\frac{\partial g_{11}}{\partial x^0}\right]=0\\
    \Gamma_{11}^1&=\frac{1}{2}g^{1p}\left[\frac{\partial g_{1p}}{\partial x^1}+\frac{\partial g_{p1}}{\partial x^1}-\frac{\partial g_{11}}{\partial x^p}\right]=\frac{1}{2}g^{11}\left[\frac{\partial g_{11}}{\partial x^1}+\frac{\partial g_{11}}{\partial x^1}-\frac{\partial g_{11}}{\partial x^1}\right]\\
    &=\frac{1}{2}g^{11}\frac{\partial g_{11}}{\partial x^1}=\frac{1}{2}\left(-e^{-2\psi}\right)\frac{\partial}{\partial u}\left(-e^{2\psi}\right)\\
    &=\psi^{\prime}\\
    \Gamma_{11}^2&=\frac{1}{2}g^{2p}\left[\frac{\partial g_{1p}}{\partial x^1}+\frac{\partial g_{p1}}{\partial x^1}-\frac{\partial g_{11}}{\partial x^p}\right]=\frac{1}{2}g^{22}\left[\frac{\partial g_{12}}{\partial x^1}+\frac{\partial g_{21}}{\partial x^1}-\frac{\partial g_{11}}{\partial x^2}\right]\\
    &=-\frac{1}{2}g^{22}\frac{\partial g_{11}}{\partial x^2}=-\frac{1}{2}\left(-e^{-2\mu_1}\right)\frac{\partial }{\partial\theta}\left(-e^{2\psi}\right)\\
    &=-e^{2(\psi-\mu_1)}\Dot{\psi}\\
    \Gamma_{11}^3&=\frac{1}{2}g^{3p}\left[\frac{\partial g_{1p}}{\partial x^1}+\frac{\partial g_{p1}}{\partial x^1}-\frac{\partial g_{11}}{\partial x^p}\right]=\frac{1}{2}g^{33}\left[\frac{\partial g_{13}}{\partial x^1}+\frac{\partial g_{31}}{\partial x^1}-\frac{\partial g_{11}}{\partial x^3}\right]=0\\
    \Gamma_{22}^0&=\frac{1}{2}g^{0p}\left[\frac{\partial g_{2p}}{\partial x^2}+\frac{\partial g_{p2}}{\partial x^2}-\frac{\partial g_{22}}{\partial x^p}\right]=\frac{1}{2}g^{00}\left[\frac{\partial g_{20}}{\partial x^2}+\frac{\partial g_{02}}{\partial x^2}-\frac{\partial g_{22}}{\partial x^0}\right]=0\\
     \Gamma_{22}^1&=\frac{1}{2}g^{1p}\left[\frac{\partial g_{2p}}{\partial x^2}+\frac{\partial g_{p2}}{\partial x^2}-\frac{\partial g_{22}}{\partial x^p}\right]=\frac{1}{2}g^{11}\left[\frac{\partial g_{21}}{\partial x^2}+\frac{\partial g_{12}}{\partial x^2}-\frac{\partial g_{22}}{\partial x^1}\right]\\
    &=-\frac{1}{2}g^{11}\frac{\partial g_{22}}{\partial x^1}=-\frac{1}{2}\left(-e^{-2\psi}\right)\frac{\partial }{\partial u}\left(-e^{2\mu_1}\right)\\
    &=-e^{2(\mu_1-\psi)}\mu_1^{\prime}
    \end{align*}
    \begin{align*}
    \Gamma_{22}^2&=\frac{1}{2}g^{2p}\left[\frac{\partial g_{2p}}{\partial x^2}+\frac{\partial g_{p2}}{\partial x^2}-\frac{\partial g_{22}}{\partial x^p}\right]=\frac{1}{2}g^{22}\left[\frac{\partial g_{22}}{\partial x^2}+\frac{\partial g_{22}}{\partial x^2}-\frac{\partial g_{22}}{\partial x^2}\right]\\
    &=\frac{1}{2}g^{22}\frac{\partial g_{22}}{\partial x^2}=\frac{1}{2}\left(-e^{-2\mu_1}\right)\frac{\partial }{\partial \theta}\left(-e^{2\mu_1}\right)\\
    &=\Dot{\mu_1}\\
    \Gamma_{22}^3&=\frac{1}{2}g^{3p}\left[\frac{\partial g_{2p}}{\partial x^2}+\frac{\partial g_{p2}}{\partial x^2}-\frac{\partial g_{22}}{\partial x^p}\right]=\frac{1}{2}g^{33}\left[\frac{\partial g_{23}}{\partial x^2}+\frac{\partial g_{32}}{\partial x^2}-\frac{\partial g_{22}}{\partial x^3}\right]=0\\
    \Gamma_{33}^0&=\frac{1}{2}g^{0p}\left[\frac{\partial g_{3p}}{\partial x^3}+\frac{\partial g_{p3}}{\partial x^3}-\frac{\partial g_{33}}{\partial x^p}\right]=\frac{1}{2}g^{00}\left[\frac{\partial g_{30}}{\partial x^3}+\frac{\partial g_{03}}{\partial x^3}-\frac{\partial g_{33}}{\partial x^0}\right]=0\\
    \Gamma_{33}^1&=\frac{1}{2}g^{1p}\left[\frac{\partial g_{3p}}{\partial x^3}+\frac{\partial g_{p3}}{\partial x^3}-\frac{\partial g_{33}}{\partial x^p}\right]=\frac{1}{2}g^{11}\left[\frac{\partial g_{31}}{\partial x^3}+\frac{\partial g_{13}}{\partial x^3}-\frac{\partial g_{33}}{\partial x^1}\right]\\
    &=-\frac{1}{2}g^{11}\frac{\partial g_{33}}{\partial x^1}=-\frac{1}{2}\left(-e^{-2\psi}\right)\frac{\partial }{\partial u}\left(-e^{2\mu_2}\right)\\
    &=-e^{2(\mu_2-\psi)}\mu_2^{\prime}\\
    \Gamma_{33}^2&=\frac{1}{2}g^{2p}\left[\frac{\partial g_{3p}}{\partial x^3}+\frac{\partial g_{p3}}{\partial x^3}-\frac{\partial g_{33}}{\partial x^p}\right]=\frac{1}{2}g^{22}\left[\frac{\partial g_{32}}{\partial x^3}+\frac{\partial g_{23}}{\partial x^3}-\frac{\partial g_{33}}{\partial x^2}\right]\\
    &=-\frac{1}{2}g^{22}\frac{\partial g_{33}}{\partial x^2}=-\frac{1}{2}\left(-e^{-2\mu_1}\right)\frac{\partial }{\partial \theta}\left(-e^{2\mu_2}\right)\\
    &=-e^{2(\mu_2-\mu_1)}\Dot{\mu_2}\\
    \Gamma_{33}^3&=\frac{1}{2}g^{3p}\left[\frac{\partial g_{3p}}{\partial x^3}+\frac{\partial g_{p3}}{\partial x^3}-\frac{\partial g_{33}}{\partial x^p}\right]=\frac{1}{2}g^{33}\left[\frac{\partial g_{33}}{\partial x^3}+\frac{\partial g_{33}}{\partial x^3}-\frac{\partial g_{33}}{\partial x^3}\right]=0\\
    \Gamma_{10}^0=\Gamma_{01}^0&=\frac{1}{2}g^{0p}\left[\frac{\partial g_{1p}}{\partial x^0}+\frac{\partial g_{p0}}{\partial x^1}-\frac{\partial g_{10}}{\partial x^p}\right]=\frac{1}{2}g^{00}\left[\frac{\partial g_{10}}{\partial x^0}+\frac{\partial g_{00}}{\partial x^1}-\frac{\partial g_{10}}{\partial x^0}\right]\\
    &=\frac{1}{2}g^{00}\frac{\partial g_{00}}{\partial x^1}=\frac{1}{2}\left(e^{-2\nu}\right)\frac{\partial }{\partial u}\left(e^{2\nu}\right)\\
    &=\nu^{\prime}\\
    \Gamma_{12}^0=\Gamma_{21}^0&=\frac{1}{2}g^{0p}\left[\frac{\partial g_{1p}}{\partial x^2}+\frac{\partial g_{p2}}{\partial x^1}-\frac{\partial g_{12}}{\partial x^p}\right]=\frac{1}{2}g^{00}\left[\frac{\partial g_{10}}{\partial x^2}+\frac{\partial g_{02}}{\partial x^1}-\frac{\partial g_{12}}{\partial x^0}\right]=0\\
    \Gamma_{13}^0=\Gamma_{31}^0&=\frac{1}{2}g^{0p}\left[\frac{\partial g_{1p}}{\partial x^3}+\frac{\partial g_{p3}}{\partial x^1}-\frac{\partial g_{13}}{\partial x^p}\right]=\frac{1}{2}g^{00}\left[\frac{\partial g_{10}}{\partial x^3}+\frac{\partial g_{03}}{\partial x^1}-\frac{\partial g_{13}}{\partial x^0}\right]=0\\
    \Gamma_{20}^0=\Gamma_{02}^0&=\frac{1}{2}g^{0p}\left[\frac{\partial g_{2p}}{\partial x^0}+\frac{\partial g_{p0}}{\partial x^2}-\frac{\partial g_{20}}{\partial x^p}\right]=\frac{1}{2}g^{00}\left[\frac{\partial g_{20}}{\partial x^0}+\frac{\partial g_{00}}{\partial x^2}-\frac{\partial g_{20}}{\partial x^0}\right]\\
    &=\frac{1}{2}g^{00}\frac{\partial g_{00}}{\partial x^2}=\frac{1}{2}\left(e^{-2\nu}\right)\frac{\partial }{\partial \theta}\left(e^{2\nu}\right)\\
    &=\Dot{\nu}\\
    \Gamma_{23}^0=\Gamma_{32}^0&=\frac{1}{2}g^{0p}\left[\frac{\partial g_{2p}}{\partial x^3}+\frac{\partial g_{p3}}{\partial x^2}-\frac{\partial g_{23}}{\partial x^p}\right]=\frac{1}{2}g^{00}\left[\frac{\partial g_{20}}{\partial x^3}+\frac{\partial g_{03}}{\partial x^2}-\frac{\partial g_{23}}{\partial x^0}\right]=0\\
    \Gamma_{30}^0=\Gamma_{03}^0&=\frac{1}{2}g^{0p}\left[\frac{\partial g_{3p}}{\partial x^0}+\frac{\partial g_{p0}}{\partial x^3}-\frac{\partial g_{30}}{\partial x^p}\right]=\frac{1}{2}g^{00}\left[\frac{\partial g_{30}}{\partial x^0}+\frac{\partial g_{00}}{\partial x^3}-\frac{\partial g_{30}}{\partial x^0}\right]=0\\
    \Gamma_{10}^1=\Gamma_{01}^1&=\frac{1}{2}g^{1p}\left[\frac{\partial g_{1p}}{\partial x^0}+\frac{\partial g_{p0}}{\partial x^1}-\frac{\partial g_{10}}{\partial x^p}\right]=\frac{1}{2}g^{11}\left[\frac{\partial g_{11}}{\partial x^0}+\frac{\partial g_{10}}{\partial x^1}-\frac{\partial g_{10}}{\partial x^1}\right]=0
    \end{align*}
    \begin{align*}
     \Gamma_{10}^2=\Gamma_{01}^2&=\frac{1}{2}g^{2p}\left[\frac{\partial g_{1p}}{\partial x^0}+\frac{\partial g_{p0}}{\partial x^1}-\frac{\partial g_{10}}{\partial x^p}\right]=\frac{1}{2}g^{22}\left[\frac{\partial g_{12}}{\partial x^0}+\frac{\partial g_{20}}{\partial x^1}-\frac{\partial g_{10}}{\partial x^2}\right]=0\\
    \Gamma_{10}^3=\Gamma_{01}^3&=\frac{1}{2}g^{3p}\left[\frac{\partial g_{1p}}{\partial x^0}+\frac{\partial g_{p0}}{\partial x^1}-\frac{\partial g_{10}}{\partial x^p}\right]=\frac{1}{2}g^{33}\left[\frac{\partial g_{13}}{\partial x^0}+\frac{\partial g_{30}}{\partial x^1}-\frac{\partial g_{10}}{\partial x^3}\right]=0\\
    \Gamma_{12}^1=\Gamma_{21}^1&=\frac{1}{2}g^{1p}\left[\frac{\partial g_{1p}}{\partial x^2}+\frac{\partial g_{p2}}{\partial x^1}-\frac{\partial g_{12}}{\partial x^p}\right]=\frac{1}{2}g^{11}\left[\frac{\partial g_{11}}{\partial x^2}+\frac{\partial g_{12}}{\partial x^1}-\frac{\partial g_{12}}{\partial x^1}\right]\\
    &=\frac{1}{2}g^{11}\frac{\partial g_{11}}{\partial x^2}=\frac{1}{2}\left(-e^{-2\psi}\right)\frac{\partial }{\partial \theta}\left(-e^{2\psi}\right)\\
    &=\Dot{\psi}\\
    \Gamma_{12}^2=\Gamma_{21}^2&=\frac{1}{2}g^{2p}\left[\frac{\partial g_{1p}}{\partial x^2}+\frac{\partial g_{p2}}{\partial x^1}-\frac{\partial g_{12}}{\partial x^p}\right]=\frac{1}{2}g^{22}\left[\frac{\partial g_{12}}{\partial x^2}+\frac{\partial g_{22}}{\partial x^1}-\frac{\partial g_{12}}{\partial x^2}\right]\\
    &=\frac{1}{2}g^{22}\frac{\partial g_{22}}{\partial x^1}=\frac{1}{2}\left(-e^{-2\mu_1}\right)\frac{\partial }{\partial u}\left(-e^{2\mu_1}\right)\\
    &=\mu_1^{\prime}\\
    \Gamma_{12}^3=\Gamma_{21}^3&=\frac{1}{2}g^{3p}\left[\frac{\partial g_{1p}}{\partial x^2}+\frac{\partial g_{p2}}{\partial x^1}-\frac{\partial g_{12}}{\partial x^p}\right]=\frac{1}{2}g^{33}\left[\frac{\partial g_{13}}{\partial x^2}+\frac{\partial g_{32}}{\partial x^1}-\frac{\partial g_{12}}{\partial x^3}\right]=0\\
    \Gamma_{13}^1=\Gamma_{31}^1&=\frac{1}{2}g^{1p}\left[\frac{\partial g_{1p}}{\partial x^3}+\frac{\partial g_{p3}}{\partial x^1}-\frac{\partial g_{13}}{\partial x^p}\right]=\frac{1}{2}g^{11}\left[\frac{\partial g_{11}}{\partial x^3}+\frac{\partial g_{13}}{\partial x^1}-\frac{\partial g_{13}}{\partial x^1}\right]=0\\
     \Gamma_{13}^2=\Gamma_{31}^2&=\frac{1}{2}g^{2p}\left[\frac{\partial g_{1p}}{\partial x^3}+\frac{\partial g_{p3}}{\partial x^1}-\frac{\partial g_{13}}{\partial x^p}\right]=\frac{1}{2}g^{22}\left[\frac{\partial g_{12}}{\partial x^3}+\frac{\partial g_{23}}{\partial x^1}-\frac{\partial g_{13}}{\partial x^2}\right]=0\\
    \Gamma_{13}^3=\Gamma_{31}^3&=\frac{1}{2}g^{3p}\left[\frac{\partial g_{1p}}{\partial x^3}+\frac{\partial g_{p3}}{\partial x^1}-\frac{\partial g_{13}}{\partial x^p}\right]=\frac{1}{2}g^{33}\left[\frac{\partial g_{13}}{\partial x^3}+\frac{\partial g_{33}}{\partial x^1}-\frac{\partial g_{13}}{\partial x^3}\right]\\
    &=\frac{1}{2}g^{33}\frac{\partial g_{33}}{\partial x^1}=\frac{1}{2}\left(-e^{-2\mu_2}\right)\frac{\partial }{\partial u}\left(-e^{2\mu_2}\right)\\
    &=\mu_2^{\prime}\\
    \Gamma_{20}^1=\Gamma_{02}^1&=\frac{1}{2}g^{1p}\left[\frac{\partial g_{2p}}{\partial x^0}+\frac{\partial g_{p0}}{\partial x^2}-\frac{\partial g_{20}}{\partial x^p}\right]=\frac{1}{2}g^{11}\left[\frac{\partial g_{21}}{\partial x^0}+\frac{\partial g_{10}}{\partial x^2}-\frac{\partial g_{20}}{\partial x^1}\right]=0\\
    \Gamma_{20}^2=\Gamma_{02}^2&=\frac{1}{2}g^{2p}\left[\frac{\partial g_{2p}}{\partial x^0}+\frac{\partial g_{p0}}{\partial x^2}-\frac{\partial g_{20}}{\partial x^p}\right]=\frac{1}{2}g^{22}\left[\frac{\partial g_{22}}{\partial x^0}+\frac{\partial g_{20}}{\partial x^2}-\frac{\partial g_{20}}{\partial x^2}\right]=0\\
    \Gamma_{20}^3=\Gamma_{02}^3&=\frac{1}{2}g^{3p}\left[\frac{\partial g_{2p}}{\partial x^0}+\frac{\partial g_{p0}}{\partial x^2}-\frac{\partial g_{20}}{\partial x^p}\right]=\frac{1}{2}g^{33}\left[\frac{\partial g_{23}}{\partial x^0}+\frac{\partial g_{30}}{\partial x^2}-\frac{\partial g_{20}}{\partial x^3}\right]=0\\
    \Gamma_{23}^1=\Gamma_{32}^1&=\frac{1}{2}g^{1p}\left[\frac{\partial g_{2p}}{\partial x^3}+\frac{\partial g_{p3}}{\partial x^2}-\frac{\partial g_{23}}{\partial x^p}\right]=\frac{1}{2}g^{11}\left[\frac{\partial g_{21}}{\partial x^3}+\frac{\partial g_{13}}{\partial x^2}-\frac{\partial g_{23}}{\partial x^1}\right]=0\\
    \Gamma_{23}^2=\Gamma_{32}^2&=\frac{1}{2}g^{2p}\left[\frac{\partial g_{2p}}{\partial x^3}+\frac{\partial g_{p3}}{\partial x^2}-\frac{\partial g_{23}}{\partial x^p}\right]=\frac{1}{2}g^{22}\left[\frac{\partial g_{22}}{\partial x^3}+\frac{\partial g_{23}}{\partial x^2}-\frac{\partial g_{23}}{\partial x^2}\right]=0\\
    \Gamma_{23}^3=\Gamma_{32}^3&=\frac{1}{2}g^{3p}\left[\frac{\partial g_{2p}}{\partial x^3}+\frac{\partial g_{p3}}{\partial x^2}-\frac{\partial g_{23}}{\partial x^p}\right]=\frac{1}{2}g^{33}\left[\frac{\partial g_{23}}{\partial x^3}+\frac{\partial g_{33}}{\partial x^2}-\frac{\partial g_{23}}{\partial x^3}\right]\\
    &=\frac{1}{2}g^{33}\frac{\partial g_{33}}{\partial x^2}=\frac{1}{2}\left(-e^{-2\mu_2}\right)\frac{\partial }{\partial \theta}\left(-e^{2\mu_2}\right)\\
    &=\Dot{\mu_2}
    \end{align*}
    \begin{align*}
    \Gamma_{30}^1=\Gamma_{03}^1&=\frac{1}{2}g^{1p}\left[\frac{\partial g_{3p}}{\partial x^0}+\frac{\partial g_{p0}}{\partial x^3}-\frac{\partial g_{30}}{\partial x^p}\right]=\frac{1}{2}g^{11}\left[\frac{\partial g_{31}}{\partial x^0}+\frac{\partial g_{10}}{\partial x^3}-\frac{\partial g_{30}}{\partial x^1}\right]=0\\
    \Gamma_{30}^2=\Gamma_{03}^2&=\frac{1}{2}g^{2p}\left[\frac{\partial g_{3p}}{\partial x^0}+\frac{\partial g_{p0}}{\partial x^3}-\frac{\partial g_{30}}{\partial x^p}\right]=\frac{1}{2}g^{22}\left[\frac{\partial g_{32}}{\partial x^0}+\frac{\partial g_{20}}{\partial x^3}-\frac{\partial g_{30}}{\partial x^2}\right]=0\\
    \Gamma_{30}^3=\Gamma_{03}^3&=\frac{1}{2}g^{3p}\left[\frac{\partial g_{3p}}{\partial x^0}+\frac{\partial g_{p0}}{\partial x^3}-\frac{\partial g_{30}}{\partial x^p}\right]=\frac{1}{2}g^{33}\left[\frac{\partial g_{33}}{\partial x^0}+\frac{\partial g_{30}}{\partial x^3}-\frac{\partial g_{30}}{\partial x^3}\right]=0
\end{align*}

\section{Ricci tensors, Ricci scalar \& Einstein tensors}\label{secA2}
\subsection{Ricci tensors}\label{subsecA2.1}
The Ricci tensors for the line element \eqref{eq:seven} are as follows:
\begin{align}
    R_{00}=&R^n_{0n0}\nonumber\\
    =&\frac{\partial \Gamma_{0n}^n}{\partial x^0}-\frac{\partial \Gamma_{00}^n}{\partial x^n}+\Gamma_{p0}^n\Gamma_{0n}^p-\Gamma_{pn}^n\Gamma_{00}^p\nonumber\\
    =&-\frac{\partial \Gamma_{00}^1}{\partial x^1}-\frac{\partial \Gamma_{00}^2}{\partial x^2}+\Gamma_{20}^0\Gamma_{00}^2+\Gamma_{00}^1\Gamma_{01}^0\nonumber\\
    &-\Gamma_{11}^1\Gamma_{00}^1-\Gamma_{21}^1\Gamma_{00}^2-\Gamma_{12}^2\Gamma_{00}^1-\Gamma_{22}^2\Gamma_{00}^2-\Gamma_{13}^3\Gamma_{00}^1-\Gamma_{23}^3\Gamma_{00}^2\nonumber\\
     =&-\frac{\partial}{\partial u} \left(e^{2(\nu-\psi)}\nu^{\prime}\right)-\frac{\partial }{\partial \theta}\left(e^{2(\nu-\mu_1)}\Dot{\nu}\right)+\left(\Dot{\nu}\right)\left(e^{2(\nu-\mu_1)}\Dot{\nu}\right)+\left(e^{2(\nu-\psi)}\nu^{\prime}\right)\left(\nu^{\prime}\right)\nonumber\\
     &-\left(\psi^{\prime}\right)\left(e^{2(\nu-\psi)}\nu^{\prime}\right)-\left(\Dot{\psi}\right)\left(e^{2(\nu-\mu_1)}\Dot{\nu}\right)-\left(\mu_1^{\prime}\right)\left(e^{2(\nu-\psi)}\nu^{\prime}\right)-\left(\Dot{\mu_1}\right)\left(e^{2(\nu-\mu_1)}\Dot{\nu}\right)\nonumber\\
    &-\left(\mu_2^{\prime}\right)\left(e^{2(\nu-\psi)}\nu^{\prime}\right)-\left(\Dot{\mu_2}\right)\left(e^{2(\nu-\mu_1)}\Dot{\nu}\right)\nonumber\\
    =& -e^{2(\nu-\psi)}\nu^{\prime\prime}-2e^{2(\nu-\psi)}\nu^{\prime}\nu^{\prime}+2e^{2(\nu-\psi)}\nu^{\prime}\psi^{\prime}+\left(e^{2(\nu-\psi)}\nu^{\prime}\right)\left(\nu^{\prime}\right)\nonumber\\
    &-\left(\psi^{\prime}\right)\left(e^{2(\nu-\psi)}\nu^{\prime}\right)-\left(\mu_1^{\prime}\right)\left(e^{2(\nu-\psi)}\nu^{\prime}\right)-\left(\mu_2^{\prime}\right)\left(e^{2(\nu-\psi)}\nu^{\prime}\right)\nonumber\\
    &-e^{2(\nu-\mu_1)}\Ddot{\nu}-2e^{2(\nu-\mu_1)}\Dot{\nu}\Dot{\nu}+2e^{2(\nu-\mu_1)}\Dot{\nu}\Dot{\mu_1}+\left(\Dot{\nu}\right)\left(e^{2(\nu-\mu_1)}\Dot{\nu}\right)\nonumber\\
    &-\left(\Dot{\psi}\right)\left(e^{2(\nu-\mu_1)}\Dot{\nu}\right)-\left(\Dot{\mu_1}\right)\left(e^{2(\nu-\mu_1)}\Dot{\nu}\right)-\left(\Dot{\mu_2}\right)\left(e^{2(\nu-\mu_1)}\Dot{\nu}\right)\nonumber\\
    = & -e^{2(\nu-\psi)}\Big[\nu^{\prime\prime}+\nu^{\prime}\left(\nu^{\prime}-\psi^{\prime}+\mu_1^{\prime}+\mu_2^{\prime}\right)\Big]-e^{2(\nu-\mu_1)}\Big[\Ddot{\nu}+\Dot{\nu}\left(\Dot{\nu}-\Dot{\mu_1}+\Dot{\psi}+\Dot{\mu_2}\right)\Big]
\end{align}

\begin{align*}
     R_{11}=&R^n_{1n1}\nonumber\\
     =&\frac{\partial \Gamma_{1n}^n}{\partial x^1}-\frac{\partial \Gamma_{11}^n}{\partial x^n}+\Gamma_{p1}^n\Gamma_{1n}^p-\Gamma_{pn}^n\Gamma_{11}^p\nonumber\\
     =&\frac{\partial \Gamma_{10}^0}{\partial x^1}+\frac{\partial \Gamma_{11}^1}{\partial x^1}+\frac{\partial \Gamma_{12}^2}{\partial x^1}+\frac{\partial \Gamma_{13}^3}{\partial x^1}-\frac{\partial \Gamma_{11}^1}{\partial x^1}-\frac{\partial \Gamma_{11}^2}{\partial x^2}\nonumber\\
     &+\Gamma_{01}^0\Gamma_{10}^0+\Gamma_{11}^0\Gamma_{10}^1+\Gamma_{21}^0\Gamma_{10}^2+\Gamma_{31}^0\Gamma_{10}^3+\Gamma_{01}^1\Gamma_{11}^0+\Gamma_{11}^1\Gamma_{11}^1\nonumber\\
     &+\Gamma_{21}^1\Gamma_{11}^2+\Gamma_{31}^1\Gamma_{11}^3+\Gamma_{01}^2\Gamma_{12}^0+\Gamma_{11}^2\Gamma_{12}^1+\Gamma_{21}^2\Gamma_{12}^2+\Gamma_{31}^2\Gamma_{12}^3+\Gamma_{01}^3\Gamma_{13}^0\nonumber\\
     &+\Gamma_{11}^3\Gamma_{13}^1+\Gamma_{21}^3\Gamma_{13}^2+\Gamma_{31}^3\Gamma_{13}^3-\Gamma_{00}^0\Gamma_{11}^0-\Gamma_{10}^0\Gamma_{11}^1-\Gamma_{20}^0\Gamma_{11}^2-\Gamma_{30}^0\Gamma_{11}^3\nonumber\\
     &-\Gamma_{01}^1\Gamma_{11}^0-\Gamma_{11}^1\Gamma_{11}^1-\Gamma_{21}^1\Gamma_{11}^2-\Gamma_{31}^1\Gamma_{11}^3-\Gamma_{02}^2\Gamma_{11}^0-\Gamma_{12}^2\Gamma_{11}^1-\Gamma_{22}^2\Gamma_{11}^2\nonumber\\
     &-\Gamma_{32}^2\Gamma_{11}^3-\Gamma_{03}^3\Gamma_{11}^0-\Gamma_{13}^3\Gamma_{11}^1-\Gamma_{23}^3\Gamma_{11}^2-\Gamma_{33}^3\Gamma_{11}^3
    \end{align*}
    \begin{align}
     =&\frac{\partial \Gamma_{10}^0}{\partial x^1}+\frac{\partial \Gamma_{12}^2}{\partial x^1}+\frac{\partial \Gamma_{13}^3}{\partial x^1}-\frac{\partial \Gamma_{11}^2}{\partial x^2}+\Gamma_{01}^0\Gamma_{10}^0+\Gamma_{11}^2\Gamma_{12}^1+\Gamma_{21}^2\Gamma_{12}^2+\Gamma_{31}^3\Gamma_{13}^3\nonumber\\
     &-\Gamma_{10}^0\Gamma_{11}^1-\Gamma_{20}^0\Gamma_{11}^2-\Gamma_{12}^2\Gamma_{11}^1-\Gamma_{22}^2\Gamma_{11}^2-\Gamma_{13}^3\Gamma_{11}^1-\Gamma_{23}^3\Gamma_{11}^2\nonumber\\
     =&\frac{\partial }{\partial u}\left(\nu^{\prime}+\mu_1^{\prime}+\mu_2^{\prime}\right)-\frac{\partial }{\partial \theta}\left(-e^{2(\psi-\mu_1)}\Dot{\psi}\right)+\left(\nu^{\prime}\right)^2+\left(-e^{2(\psi-\mu_1)}\Dot{\psi}\right)\left(\Dot{\psi}\right)\nonumber\\
     &+\left(\mu_1^{\prime}\right)^2+\left(\mu_2^{\prime}\right)^2-\left(\nu^{\prime}\right)\left(\psi^{\prime}
\right)-\left(\Dot{\nu}\right)\left(-e^{2(\psi-\mu_1)}\Dot{\psi}\right)-\left(\mu_1^{\prime}\right)\left(\psi^{\prime}\right)\nonumber\\
&-\left(\Dot{\mu_1}\right)\left(-e^{2(\psi-\mu_1)}\Dot{\psi}\right)-\left(\mu_2^{\prime}\right)\left(\psi^{\prime}\right)-\left(\Dot{\mu_2}\right)\left(-e^{2(\psi-\mu_1)}\Dot{\psi}\right)\nonumber\\
=&\left(\nu^{\prime\prime}+\mu_1^{\prime\prime}+\mu_2^{\prime\prime}\right)+\left(\nu^{\prime}\right)^2+\left(\mu_1^{\prime}\right)^2+\left(\mu_2^{\prime}\right)^2-\left(\nu^{\prime}\right)\left(\psi^{\prime}\right)-\left(\mu_1^{\prime}\right)\left(\psi^{\prime}\right)-\left(\mu_2^{\prime}\right)\left(\psi^{\prime}\right)\nonumber\\
    &+e^{2(\psi-\mu_1)}\Ddot{\psi}+2e^{2(\psi-\mu_1)}\Dot{\psi}\Dot{\psi}-2e^{2(\psi-\mu_1)}\Dot{\psi}\Dot{\mu_1}+\left(-e^{2(\psi-\mu_1)}\Dot{\psi}\right)\left(\Dot{\psi}\right)\nonumber\\
    &-\left(\Dot{\nu}\right)\left(-e^{2(\psi-\mu_1)}\Dot{\psi}\right)-\left(\Dot{\mu_1}\right)\left(-e^{2(\psi-\mu_1)}\Dot{\psi}\right)-\left(\Dot{\mu_2}\right)\left(-e^{2(\psi-\mu_1)}\Dot{\psi}\right)\nonumber\\
    &\nonumber\\
    =&\nu^{\prime\prime}+\mu_1^{\prime\prime}+\mu_2^{\prime\prime}+\left(\nu^{\prime}\right)^2+\left(\mu_1^{\prime}\right)^2+\left(\mu_2^{\prime}\right)^2-\nu^{\prime}\psi^{\prime}-\mu_1^{\prime}\psi^{\prime}-\mu_2^{\prime}\psi^{\prime}\nonumber\\
    &+e^{2(\psi-\mu_1)}\Big(\Ddot{\psi}+\Dot{\psi}\Dot{\psi}-\Dot{\psi}\Dot{\mu_1}+\Dot{\nu}\Dot{\psi}+\Dot{\mu_2}\Dot{\psi}\Big)
\end{align}

\begin{align}
     R_{22}=&R^n_{2n2}\nonumber\\
     =&\frac{\partial \Gamma_{2n}^n}{\partial x^2}-\frac{\partial \Gamma_{22}^n}{\partial x^n}+\Gamma_{p2}^n\Gamma_{2n}^p-\Gamma_{pn}^n\Gamma_{22}^p\nonumber\\
     =&\frac{\partial \Gamma_{20}^0}{\partial x^2}+\frac{\partial \Gamma_{21}^1}{\partial x^2}+\frac{\partial \Gamma_{22}^2}{\partial x^2}+\frac{\partial \Gamma_{23}^3}{\partial x^2}-\frac{\partial \Gamma_{22}^1}{\partial x^1}-\frac{\partial \Gamma_{22}^2}{\partial x^2}\nonumber\\
     &+\Gamma_{p2}^0\Gamma_{20}^p+\Gamma_{p2}^1\Gamma_{21}^p+\Gamma_{p2}^2\Gamma_{22}^p+\Gamma_{p2}^3\Gamma_{23}^p-\Gamma_{p0}^0\Gamma_{22}^p-\Gamma_{p1}^1\Gamma_{22}^p\nonumber\\
     &-\Gamma_{p2}^2\Gamma_{22}^p-\Gamma_{p3}^3\Gamma_{22}^p\nonumber\\
      =&\frac{\partial \Gamma_{20}^0}{\partial x^2}+\frac{\partial \Gamma_{21}^1}{\partial x^2}+\frac{\partial \Gamma_{23}^3}{\partial x^2}-\frac{\partial \Gamma_{22}^1}{\partial x^1}+\Gamma_{02}^0\Gamma_{20}^0+\Gamma_{12}^0\Gamma_{20}^1+\Gamma_{22}^0\Gamma_{20}^2+\Gamma_{32}^0\Gamma_{20}^3\nonumber\\
      &+\Gamma_{02}^1\Gamma_{21}^0+\Gamma_{12}^1\Gamma_{21}^1+\Gamma_{22}^1\Gamma_{21}^2+\Gamma_{32}^1\Gamma_{21}^3+\Gamma_{02}^2\Gamma_{22}^0+\Gamma_{12}^2\Gamma_{22}^1+\Gamma_{22}^2\Gamma_{22}^2\nonumber\\
      &+\Gamma_{32}^2\Gamma_{22}^3+\Gamma_{02}^3\Gamma_{23}^0+\Gamma_{12}^3\Gamma_{23}^1+\Gamma_{22}^3\Gamma_{23}^2+\Gamma_{32}^3\Gamma_{23}^3-\Gamma_{00}^0\Gamma_{22}^0-\Gamma_{10}^0\Gamma_{22}^1\nonumber\\
      &-\Gamma_{20}^0\Gamma_{22}^2-\Gamma_{30}^0\Gamma_{22}^3-\Gamma_{01}^1\Gamma_{22}^0-\Gamma_{11}^1\Gamma_{22}^1-\Gamma_{21}^1\Gamma_{22}^2-\Gamma_{31}^1\Gamma_{22}^3-\Gamma_{02}^2\Gamma_{22}^0\nonumber\\
     &-\Gamma_{12}^2\Gamma_{22}^1-\Gamma_{22}^2\Gamma_{22}^2-\Gamma_{32}^2\Gamma_{22}^3-\Gamma_{03}^3\Gamma_{22}^0-\Gamma_{13}^3\Gamma_{22}^1-\Gamma_{23}^3\Gamma_{22}^2-\Gamma_{33}^3\Gamma_{22}^3\nonumber\\
     =&\Ddot{\nu}+\Ddot{\psi}+\Ddot{\mu_2}+\left(e^{2(\mu_1-\psi)}\mu_1^{\prime\prime}+2e^{2(\mu_1-\psi)}\mu_1^{\prime}(\mu_1^{\prime}-\psi^{\prime})\right)+\left(\Dot{\nu}\right)^2+\left(\Dot{\psi}\right)^2\nonumber\\
    &+2\left(-e^{2(\mu_1-\psi)}\mu_1^{\prime}\right)\left(\mu_1^{\prime}\right)+\left(\Dot{\mu_1}\right)^2+\left(\Dot{\mu_2}\right)^2-\left(\nu^{\prime}\right)\left(-e^{2(\mu_1-\psi)}\mu_1^{\prime}\right)-\left(\Dot{\nu}\right)\left(\Dot{\mu_1}\right)\nonumber\\
     &-\left(\psi^{\prime}\right)\left(-e^{2(\mu_1-\psi)}\mu_1^{\prime}\right)-\left(\Dot{\psi}\right)\left(\Dot{\mu_1}\right)-\left(\mu_1^{\prime}\right)\left(-e^{2(\mu_1-\psi)}\mu_1^{\prime}\right)-\left(\Dot{\mu_1}\right)^2\nonumber\\
     &-\left(\mu_2^{\prime}\right)\left(-e^{2(\mu_1-\psi)}\mu_1^{\prime}\right)-\left(\Dot{\mu_2}\right)\left(\Dot{\mu_1}\right)\nonumber\\
     &\nonumber\\
     =&e^{2(\mu_1-\psi)}\Big[\mu_1^{\prime\prime}-\mu_1^{\prime}\psi^{\prime}+\mu_1^{\prime}\mu_1^{\prime}+\mu_2^{\prime}\mu_1^{\prime}+\nu^{\prime}\mu_1^{\prime}\Big]-\left(\Dot{\nu}\right)\left(\Dot{\mu_1}\right)+\Ddot{\nu}\nonumber\nonumber\\
     &+\Ddot{\psi}+\Ddot{\mu_2}+\left(\Dot{\mu_2}\right)^2+\left(\Dot{\nu}\right)^2+\left(\Dot{\psi}\right)^2-\left(\Dot{\mu_2}\right)\left(\Dot{\mu_1}\right)-\left(\Dot{\psi}\right)\left(\Dot{\mu_1}\right)
\end{align}

\begin{align}
    R_{33}=&R^n_{3n3}\nonumber\\
    =&\frac{\partial \Gamma_{3n}^n}{\partial x^3}-\frac{\partial \Gamma_{33}^n}{\partial x^n}+\Gamma_{p3}^n\Gamma_{3n}^p-\Gamma_{pn}^n\Gamma_{33}^p\nonumber\\
    =&-\frac{\partial \Gamma_{33}^1}{\partial x^1}-\frac{\partial \Gamma_{33}^2}{\partial x^2}+\Gamma_{03}^0\Gamma_{30}^0+\Gamma_{13}^0\Gamma_{30}^1+\Gamma_{23}^0\Gamma_{30}^2+\Gamma_{33}^0\Gamma_{30}^3\nonumber\\
    &+\Gamma_{03}^1\Gamma_{31}^0+\Gamma_{13}^1\Gamma_{31}^1+\Gamma_{23}^1\Gamma_{31}^2+\Gamma_{33}^1\Gamma_{31}^3+\Gamma_{03}^2\Gamma_{32}^0+\Gamma_{13}^2\Gamma_{32}^1\nonumber\\
    &+\Gamma_{23}^2\Gamma_{32}^2+\Gamma_{33}^2\Gamma_{32}^3+\Gamma_{03}^3\Gamma_{33}^0+\Gamma_{13}^3\Gamma_{33}^1+\Gamma_{23}^3\Gamma_{33}^2+\Gamma_{33}^3\Gamma_{33}^3\nonumber\\
    &-\Gamma_{00}^0\Gamma_{33}^0-\Gamma_{10}^0\Gamma_{33}^1-\Gamma_{20}^0\Gamma_{33}^2-\Gamma_{30}^0\Gamma_{33}^3-\Gamma_{01}^1\Gamma_{33}^0-\Gamma_{11}^1\Gamma_{33}^1\nonumber\\
    &-\Gamma_{21}^1\Gamma_{33}^2-\Gamma_{31}^1\Gamma_{33}^3-\Gamma_{02}^2\Gamma_{33}^0-\Gamma_{12}^2\Gamma_{33}^1-\Gamma_{22}^2\Gamma_{33}^2-\Gamma_{32}^2\Gamma_{33}^3\nonumber\\
    &-\Gamma_{03}^3\Gamma_{33}^0-\Gamma_{13}^3\Gamma_{33}^1-\Gamma_{23}^3\Gamma_{33}^2-\Gamma_{33}^3\Gamma_{33}^3\nonumber\\
    =&-\frac{\partial \Gamma_{33}^1}{\partial x^1}-\frac{\partial \Gamma_{33}^2}{\partial x^2}+\Gamma_{33}^1\Gamma_{31}^3+\Gamma_{33}^2\Gamma_{32}^3+\Gamma_{13}^3\Gamma_{33}^1+\Gamma_{23}^3\Gamma_{33}^2\nonumber\\
    &-\Gamma_{10}^0\Gamma_{33}^1-\Gamma_{20}^0\Gamma_{33}^2-\Gamma_{11}^1\Gamma_{33}^1-\Gamma_{21}^1\Gamma_{33}^2-\Gamma_{12}^2\Gamma_{33}^1-\Gamma_{22}^2\Gamma_{33}^2\nonumber\\
    &-\Gamma_{13}^3\Gamma_{33}^1-\Gamma_{23}^3\Gamma_{33}^2\nonumber\\
    =&-\frac{\partial }{\partial u}\left(-e^{2(\mu_2-\psi)}\mu_2^{\prime}\right)-\frac{\partial }{\partial \theta}\left(-e^{2(\mu_2-\mu_1)}\Dot{\mu_2}\right)+\left(-e^{2(\mu_2-\psi)}\mu_2^{\prime}\right)\left(\mu_2^{\prime}\right)\nonumber\\
    &+\left(-e^{2(\mu_2-\mu_1)}\Dot{\mu_2}\right)\left(\Dot{\mu_2}\right)+\left(\mu_2^{\prime}\right)\left(-e^{2(\mu_2-\psi)}\mu_2^{\prime}\right)+\left(\Dot{\mu_2}\right)\left(-e^{2(\mu_2-\mu_1)}\Dot{\mu_2}\right)\nonumber\\
    &-\left(\nu^{\prime}\right)\left(-e^{2(\mu_2-\psi)}\mu_2^{\prime}\right)-\left(\Dot{\nu}\right)\left(-e^{2(\mu_2-\mu_1)}\Dot{\mu_2}\right)-\left(\psi^{\prime}\right)\left(-e^{2(\mu_2-\psi)}\mu_2^{\prime}\right)\nonumber\\
    &-\left(\Dot{\psi}\right)\left(-e^{2(\mu_2-\mu_1)}\Dot{\mu_2}\right)-\left(\mu_1^{\prime}\right)\left(-e^{2(\mu_2-\psi)}\mu_2^{\prime}\right)-\left(\Dot{\mu_1}\right)\left(-e^{2(\mu_2-\mu_1)}\Dot{\mu_2}\right)\nonumber\\
    &-\left(\mu_2^{\prime}\right)\left(-e^{2(\mu_2-\psi)}\mu_2^{\prime}
\right)-\left(\Dot{\mu_2}\right)\left(-e^{2(\mu_2-\mu_1)}\Dot{\mu_2}\right)\nonumber\\
=&e^{2(\mu_2-\psi)}\mu_2^{\prime\prime}+2e^{2(\mu_2-\psi)}\mu_2^{\prime}\mu_2^{\prime}-2e^{2(\mu_2-\psi)}\mu_2^{\prime}\psi^{\prime}+\left(-e^{2(\mu_2-\psi)}\mu_2^{\prime}\right)\left(\mu_2^{\prime}\right)\nonumber\\
    &-\left(\nu^{\prime}\right)\left(-e^{2(\mu_2-\psi)}\mu_2^{\prime}\right)+\left(\mu_2^{\prime}\right)\left(-e^{2(\mu_2-\psi)}\mu_2^{\prime}\right)-\left(\mu_1^{\prime}\right)\left(-e^{2(\mu_2-\psi)}\mu_2^{\prime}\right)\nonumber\\
    &-\left(\mu_2^{\prime}\right)\left(-e^{2(\mu_2-\psi)}\mu_2^{\prime}\right)-\left(\psi^{\prime}\right)\left(-e^{2(\mu_2-\psi)}\mu_2^{\prime}\right)+e^{2(\mu_2-\mu_1)}\Ddot{\mu_2}\nonumber\\
    &+2e^{2(\mu_2-\mu_1)}\Dot{\mu_2}\Dot{\mu_2}-2e^{2(\mu_2-\mu_1)}\Dot{\mu_2}\Dot{\mu_1}-\left(\Dot{\nu}\right)\left(-e^{2(\mu_2-\mu_1)}\Dot{\mu_2}\right)\nonumber\\
    &-\left(\Dot{\psi}\right)\left(-e^{2(\mu_2-\mu_1)}\Dot{\mu_2}\right)-\left(\Dot{\mu_1}\right)\left(-e^{2(\mu_2-\mu_1)}\Dot{\mu_2}\right)-\left(\Dot{\mu_2}\right)\left(-e^{2(\mu_2-\mu_1)}\Dot{\mu_2}\right)\nonumber\\
    &+\left(-e^{2(\mu_2-\mu_1)}\Dot{\mu_2}\right)\left(\Dot{\mu_2}\right)+\left(\Dot{\mu_2}\right)\left(-e^{2(\mu_2-\mu_1)}\Dot{\mu_2}\right)\nonumber\\
    &\nonumber\\
    =&e^{2(\mu_2-\psi)}\Big[\mu_2^{\prime\prime}+\mu_2^{\prime}\left(\mu_2^{\prime}-\psi^{\prime}+\nu^{\prime}+\mu_1^{\prime}\right)\Big]\nonumber\\
    &\hspace{3cm}+e^{2(\mu_2-\mu_1)}\Big[\Ddot{\mu_2}+\Dot{\mu_2}\left(\Dot{\mu_2}-\Dot{\mu_1}+\Dot{\nu}+\Dot{\psi}\right)\Big]
\end{align}

\begin{align}
    R_{12}=&R^n_{1n2}\nonumber\\
    =&\frac{\partial \Gamma_{1n}^n}{\partial x^2}-\frac{\partial \Gamma_{12}^n}{\partial x^n}+\Gamma_{p2}^n\Gamma_{1n}^p-\Gamma_{pn}^n\Gamma_{12}^p\nonumber\\
    =&\frac{\partial \Gamma_{10}^0}{\partial x^2}+\frac{\partial \Gamma_{11}^1}{\partial x^2}+\frac{\partial \Gamma_{12}^2}{\partial x^2}+\frac{\partial \Gamma_{13}^3}{\partial x^2}-\frac{\partial \Gamma_{12}^1}{\partial x^1}-\frac{\partial \Gamma_{12}^2}{\partial x^2}\nonumber\\
    &+\Gamma_{p2}^0\Gamma_{10}^p+\Gamma_{p2}^1\Gamma_{11}^p+\Gamma_{p2}^2\Gamma_{12}^p+\Gamma_{p2}^3\Gamma_{13}^p\nonumber\\
    &-\Gamma_{p0}^0\Gamma_{12}^p-\Gamma_{p1}^1\Gamma_{12}^p-\Gamma_{p2}^2\Gamma_{12}^p-\Gamma_{p3}^3\Gamma_{12}^p\nonumber\\
    =&\frac{\partial \Gamma_{10}^0}{\partial x^2}+\frac{\partial \Gamma_{11}^1}{\partial x^2}+\frac{\partial \Gamma_{12}^2}{\partial x^2}+\frac{\partial \Gamma_{13}^3}{\partial x^2}-\frac{\partial \Gamma_{12}^1}{\partial x^1}-\frac{\partial \Gamma_{12}^2}{\partial x^2}\nonumber\\
    &+\Gamma_{02}^0\Gamma_{10}^0+\Gamma_{12}^0\Gamma_{10}^1+\Gamma_{22}^0\Gamma_{10}^2+\Gamma_{32}^0\Gamma_{10}^3+\Gamma_{02}^1\Gamma_{11}^0+\Gamma_{12}^1\Gamma_{11}^1\nonumber\\
    &+\Gamma_{22}^1\Gamma_{11}^2+\Gamma_{32}^1\Gamma_{11}^3+\Gamma_{02}^2\Gamma_{12}^0+\Gamma_{12}^2\Gamma_{12}^1+\Gamma_{22}^2\Gamma_{12}^2+\Gamma_{32}^2\Gamma_{12}^3\nonumber\\
    &+\Gamma_{02}^3\Gamma_{13}^0+\Gamma_{12}^3\Gamma_{13}^1+\Gamma_{22}^3\Gamma_{13}^2+\Gamma_{32}^3\Gamma_{13}^3-\Gamma_{00}^0\Gamma_{12}^0-\Gamma_{10}^0\Gamma_{12}^1\nonumber\\
    &-\Gamma_{20}^0\Gamma_{12}^2-\Gamma_{30}^0\Gamma_{12}^3-\Gamma_{01}^1\Gamma_{12}^0-\Gamma_{11}^1\Gamma_{12}^1-\Gamma_{21}^1\Gamma_{12}^2-\Gamma_{31}^1\Gamma_{12}^3\nonumber\\
    &-\Gamma_{02}^2\Gamma_{12}^0-\Gamma_{12}^2\Gamma_{12}^1-\Gamma_{22}^2\Gamma_{12}^2-\Gamma_{32}^2\Gamma_{12}^3-\Gamma_{03}^3\Gamma_{12}^0-\Gamma_{13}^3\Gamma_{12}^1\nonumber\\
    &-\Gamma_{23}^3\Gamma_{12}^2-\Gamma_{33}^3\Gamma_{12}^3\nonumber\\
     =&\frac{\partial \Gamma_{10}^0}{\partial x^2}+\frac{\partial \Gamma_{11}^1}{\partial x^2}+\frac{\partial \Gamma_{13}^3}{\partial x^2}-\frac{\partial \Gamma_{12}^1}{\partial x^1}+\Gamma_{02}^0\Gamma_{10}^0+\Gamma_{22}^1\Gamma_{11}^2\nonumber\\
    &+\Gamma_{32}^3\Gamma_{13}^3-\Gamma_{10}^0\Gamma_{12}^1-\Gamma_{20}^0\Gamma_{12}^2-\Gamma_{12}^2\Gamma_{12}^1-\Gamma_{13}^3\Gamma_{12}^1-\Gamma_{23}^3\Gamma_{12}^2\nonumber\\
    =&\frac{\partial }{\partial \theta}\left(\nu^{\prime}+\psi^{\prime}+\mu_2^{\prime}\right)-\frac{\partial }{\partial u}\left(\Dot{\psi}\right)+\left(\Dot{\nu}
\right)\left(\nu^{\prime}\right)+\left(-e^{2(\mu_1-\psi)}\mu_1^{\prime}\right)\left(-e^{2(\psi-\mu_1)}\Dot{\psi}\right)\nonumber\\
&+\left(\Dot{\mu_2}\right)\left(\mu_2^{\prime}\right)-\left(\nu^{\prime}\right)\left(\Dot{\psi}\right)-\left(\Dot{\nu}\right)\left(\mu_1^{\prime}\right)-\left(\mu_1^{\prime}\right)\left(\Dot{\psi}\right)-\left(\mu_2^{\prime}\right)\left(\Dot{\psi}\right)-\left(\Dot{\mu_2}\right)\left(\mu_1^{\prime}\right)\nonumber\\
&=\frac{\partial }{\partial \theta}\left(\nu^{\prime}+\psi^{\prime}+\mu_2^{\prime}\right)-\frac{\partial }{\partial u}\left(\Dot{\psi}\right)+\left(\Dot{\nu}
\right)\left(\nu^{\prime}\right)+\left(-e^{2(\mu_1-\psi)}\mu_1^{\prime}\right)\left(-e^{2(\psi-\mu_1)}\Dot{\psi}\right)\nonumber\\
&+\left(\Dot{\mu_2}\right)\left(\mu_2^{\prime}\right)-\left(\nu^{\prime}\right)\left(\Dot{\psi}\right)-\left(\Dot{\nu}\right)\left(\mu_1^{\prime}\right)-\left(\mu_1^{\prime}\right)\left(\Dot{\psi}\right)-\left(\mu_2^{\prime}\right)\left(\Dot{\psi}\right)-\left(\Dot{\mu_2}\right)\left(\mu_1^{\prime}\right)\nonumber\\
&\nonumber\\
=&\frac{\partial }{\partial \theta}\left(\nu^{\prime}+\mu_2^{\prime}\right)+\left(\Dot{\nu}
\right)\left(\nu^{\prime}\right)+\left(\Dot{\mu_2}\right)\left(\mu_2^{\prime}\right)-\left(\nu^{\prime}\right)\left(\Dot{\psi}\right)-\left(\Dot{\nu}\right)\left(\mu_1^{\prime}\right)\nonumber\\
&-\left(\mu_2^{\prime}\right)\left(\Dot{\psi}\right)-\left(\Dot{\mu_2}\right)\left(\mu_1^{\prime}\right)
\end{align}
The vanishing Ricci tensors are below:
\begin{align*}
    R_{01}=R^n_{0n1}&=\frac{\partial \Gamma_{0n}^n}{\partial x^1}-\frac{\partial \Gamma_{01}^n}{\partial x^n}+\Gamma_{p1}^n\Gamma_{0n}^p-\Gamma_{pn}^n\Gamma_{01}^p\\
    &=\frac{\partial \Gamma_{00}^0}{\partial x^1}+\frac{\partial \Gamma_{01}^1}{\partial x^1}+\frac{\partial \Gamma_{02}^2}{\partial x^1}+\frac{\partial \Gamma_{03}^3}{\partial x^1}-\frac{\partial \Gamma_{01}^1}{\partial x^1}-\frac{\partial \Gamma_{01}^2}{\partial x^2}\\
    &+\Gamma_{p1}^n\Gamma_{0n}^p-\Gamma_{pn}^n\Gamma_{01}^p\\
    &=\Gamma_{p1}^0\Gamma_{00}^p+\Gamma_{p1}^1\Gamma_{01}^p+\Gamma_{p1}^2\Gamma_{02}^p+\Gamma_{p1}^3\Gamma_{03}^p-\Gamma_{p0}^0\Gamma_{01}^p-\Gamma_{p1}^1\Gamma_{01}^p-\Gamma_{p2}^2\Gamma_{01}^p-\Gamma_{p3}^3\Gamma_{01}^p\\
    &=\Gamma_{01}^0\Gamma_{00}^0+\Gamma_{11}^0\Gamma_{00}^1+\Gamma_{21}^0\Gamma_{00}^2+\Gamma_{31}^0\Gamma_{00}^3+\Gamma_{01}^1\Gamma_{01}^0+\Gamma_{11}^1\Gamma_{01}^1+\Gamma_{21}^1\Gamma_{01}^2+\Gamma_{31}^1\Gamma_{01}^3\\
    &+\Gamma_{01}^2\Gamma_{02}^0+\Gamma_{11}^2\Gamma_{02}^1+\Gamma_{21}^2\Gamma_{02}^2+\Gamma_{31}^2\Gamma_{02}^3+\Gamma_{01}^3\Gamma_{03}^0+\Gamma_{11}^3\Gamma_{03}^1+\Gamma_{21}^3\Gamma_{03}^2+\Gamma_{31}^3\Gamma_{03}^3\\
    &-\Gamma_{00}^0\Gamma_{01}^0-\Gamma_{10}^0\Gamma_{01}^1-\Gamma_{20}^0\Gamma_{01}^2-\Gamma_{30}^0\Gamma_{01}^3-\Gamma_{01}^1\Gamma_{01}^0-\Gamma_{11}^1\Gamma_{01}^1-\Gamma_{21}^1\Gamma_{01}^2-\Gamma_{31}^1\Gamma_{01}^3\\
    &-\Gamma_{02}^2\Gamma_{01}^0-\Gamma_{12}^2\Gamma_{01}^1-\Gamma_{22}^2\Gamma_{01}^2-\Gamma_{32}^2\Gamma_{01}^3-\Gamma_{03}^3\Gamma_{01}^0-\Gamma_{13}^3\Gamma_{01}^1-\Gamma_{23}^3\Gamma_{01}^2-\Gamma_{33}^3\Gamma_{01}^3\\
    &=0
\end{align*}

\begin{align*}
    R_{02}=R^n_{0n2}&=\frac{\partial \Gamma_{0n}^n}{\partial x^2}-\frac{\partial \Gamma_{02}^n}{\partial x^n}+\Gamma_{p2}^n\Gamma_{0n}^p-\Gamma_{pn}^n\Gamma_{02}^p\\
    &=\frac{\partial \Gamma_{00}^0}{\partial x^2}+\frac{\partial \Gamma_{01}^1}{\partial x^2}+\frac{\partial \Gamma_{02}^2}{\partial x^2}+\frac{\partial \Gamma_{03}^3}{\partial x^2}-\frac{\partial \Gamma_{02}^1}{\partial x^1}-\frac{\partial \Gamma_{02}^2}{\partial x^2}\\
    &+\Gamma_{p2}^0\Gamma_{00}^p+\Gamma_{p2}^1\Gamma_{01}^p+\Gamma_{p2}^2\Gamma_{02}^p+\Gamma_{p2}^3\Gamma_{03}^p\\
    &-\Gamma_{p0}^0\Gamma_{02}^p-\Gamma_{p1}^1\Gamma_{02}^p-\Gamma_{p2}^2\Gamma_{02}^p-\Gamma_{p3}^3\Gamma_{02}^p\\
     &=\Gamma_{02}^0\Gamma_{00}^0+\Gamma_{12}^0\Gamma_{00}^1+\Gamma_{22}^0\Gamma_{00}^2+\Gamma_{32}^0\Gamma_{00}^3+\Gamma_{02}^1\Gamma_{01}^0+\Gamma_{12}^1\Gamma_{01}^1\\
     &+\Gamma_{22}^1\Gamma_{01}^2+\Gamma_{32}^1\Gamma_{01}^3+\Gamma_{02}^2\Gamma_{02}^0+\Gamma_{12}^2\Gamma_{02}^1+\Gamma_{22}^2\Gamma_{02}^2+\Gamma_{32}^2\Gamma_{02}^3\\
     &+\Gamma_{02}^3\Gamma_{03}^0+\Gamma_{12}^3\Gamma_{03}^1+\Gamma_{22}^3\Gamma_{03}^2+\Gamma_{32}^3\Gamma_{03}^3-\Gamma_{00}^0\Gamma_{02}^0-\Gamma_{10}^0\Gamma_{02}^1\\
     &-\Gamma_{20}^0\Gamma_{02}^2-\Gamma_{30}^0\Gamma_{02}^3-\Gamma_{01}^1\Gamma_{02}^0-\Gamma_{11}^1\Gamma_{02}^1-\Gamma_{21}^1\Gamma_{02}^2-\Gamma_{31}^1\Gamma_{02}^3\\
    &-\Gamma_{02}^2\Gamma_{02}^0-\Gamma_{12}^2\Gamma_{02}^1-\Gamma_{22}^2\Gamma_{02}^2-\Gamma_{32}^2\Gamma_{02}^3-\Gamma_{03}^3\Gamma_{02}^0-\Gamma_{13}^3\Gamma_{02}^1\\
    &-\Gamma_{23}^3\Gamma_{02}^2-\Gamma_{33}^3\Gamma_{02}^3\\
    &=0
\end{align*}
\begin{align*}
   R_{03}=R^n_{0n3}&=\frac{\partial \Gamma_{0n}^n}{\partial x^3}-\frac{\partial \Gamma_{03}^n}{\partial x^n}+\Gamma_{p3}^n\Gamma_{0n}^p-\Gamma_{pn}^n\Gamma_{03}^p\\
   &=-\frac{\partial \Gamma_{03}^1}{\partial x^1}-\frac{\partial \Gamma_{03}^2}{\partial x^2}+\Gamma_{03}^0\Gamma_{00}^0+\Gamma_{13}^0\Gamma_{00}^1+\Gamma_{23}^0\Gamma_{00}^2+\Gamma_{33}^0\Gamma_{00}^3\\
   &+\Gamma_{03}^1\Gamma_{01}^0+\Gamma_{13}^1\Gamma_{01}^1+\Gamma_{23}^1\Gamma_{01}^2+\Gamma_{33}^1\Gamma_{01}^3\\
   &+\Gamma_{03}^2\Gamma_{02}^0+\Gamma_{13}^2\Gamma_{02}^1+\Gamma_{23}^2\Gamma_{02}^2+\Gamma_{33}^2\Gamma_{02}^3\\
   &+\Gamma_{03}^3\Gamma_{03}^0+\Gamma_{13}^3\Gamma_{03}^1+\Gamma_{23}^3\Gamma_{03}^2+\Gamma_{33}^3\Gamma_{03}^3\\
   &-\Gamma_{00}^0\Gamma_{03}^0-\Gamma_{10}^0\Gamma_{03}^1-\Gamma_{20}^0\Gamma_{03}^2-\Gamma_{30}^0\Gamma_{03}^3\\
   &-\Gamma_{01}^1\Gamma_{03}^0-\Gamma_{11}^1\Gamma_{03}^1-\Gamma_{21}^1\Gamma_{03}^2-\Gamma_{31}^1\Gamma_{03}^3\\
   &-\Gamma_{02}^2\Gamma_{03}^0-\Gamma_{12}^2\Gamma_{03}^1-\Gamma_{22}^2\Gamma_{03}^2-\Gamma_{32}^2\Gamma_{03}^3\\
   &-\Gamma_{03}^3\Gamma_{03}^0-\Gamma_{13}^3\Gamma_{03}^1-\Gamma_{23}^3\Gamma_{03}^2-\Gamma_{33}^3\Gamma_{03}^3=0
\end{align*}

\begin{align*}
    R_{13}=R^n_{1n3}&=\frac{\partial \Gamma_{1n}^n}{\partial x^3}-\frac{\partial \Gamma_{13}^n}{\partial x^n}+\Gamma_{p3}^n\Gamma_{1n}^p-\Gamma_{pn}^n\Gamma_{13}^p\\
    &=-\frac{\partial \Gamma_{13}^1}{\partial x^1}-\frac{\partial \Gamma_{13}^2}{\partial x^2}+\Gamma_{p3}^0\Gamma_{10}^p+\Gamma_{p3}^1\Gamma_{11}^p+\Gamma_{p3}^2\Gamma_{12}^p+\Gamma_{p3}^3\Gamma_{13}^p\\
    &-\Gamma_{p0}^0\Gamma_{13}^p-\Gamma_{p1}^1\Gamma_{13}^p-\Gamma_{p2}^2\Gamma_{13}^p-\Gamma_{p3}^3\Gamma_{13}^p\\
    &=\Gamma_{03}^0\Gamma_{10}^0+\Gamma_{13}^0\Gamma_{10}^1+\Gamma_{23}^0\Gamma_{10}^2+\Gamma_{33}^0\Gamma_{10}^3+\Gamma_{03}^1\Gamma_{11}^0+\Gamma_{13}^1\Gamma_{11}^1\\
    &+\Gamma_{23}^1\Gamma_{11}^2+\Gamma_{33}^1\Gamma_{11}^3+\Gamma_{03}^2\Gamma_{12}^0+\Gamma_{13}^2\Gamma_{12}^1+\Gamma_{23}^2\Gamma_{12}^2+\Gamma_{33}^2\Gamma_{12}^3\\
    &+\Gamma_{03}^3\Gamma_{13}^0+\Gamma_{13}^3\Gamma_{13}^1+\Gamma_{23}^3\Gamma_{13}^2+\Gamma_{33}^3\Gamma_{13}^3-\Gamma_{00}^0\Gamma_{13}^0-\Gamma_{10}^0\Gamma_{13}^1\\
    &-\Gamma_{20}^0\Gamma_{13}^2-\Gamma_{30}^0\Gamma_{13}^3-\Gamma_{01}^1\Gamma_{13}^0-\Gamma_{11}^1\Gamma_{13}^1-\Gamma_{21}^1\Gamma_{13}^2-\Gamma_{31}^1\Gamma_{13}^3\\
    &-\Gamma_{02}^2\Gamma_{13}^0-\Gamma_{12}^2\Gamma_{13}^1-\Gamma_{22}^2\Gamma_{13}^2-\Gamma_{32}^2\Gamma_{13}^3-\Gamma_{03}^3\Gamma_{13}^0-\Gamma_{13}^3\Gamma_{13}^1\\
    &-\Gamma_{23}^3\Gamma_{13}^2-\Gamma_{33}^3\Gamma_{13}^3\\
    &=0
\end{align*}

\begin{align*}
    R_{23}=R^n_{2n3}&=\frac{\partial \Gamma_{2n}^n}{\partial x^3}-\frac{\partial \Gamma_{23}^n}{\partial x^n}+\Gamma_{p3}^n\Gamma_{2n}^p-\Gamma_{pn}^n\Gamma_{23}^p\\
    &=-\frac{\partial \Gamma_{23}^1}{\partial x^1}-\frac{\partial \Gamma_{23}^2}{\partial x^2}+\Gamma_{p3}^0\Gamma_{20}^p+\Gamma_{p3}^1\Gamma_{21}^p+\Gamma_{p3}^2\Gamma_{22}^p+\Gamma_{p3}^3\Gamma_{23}^p\\
    &-\Gamma_{p0}^0\Gamma_{23}^p-\Gamma_{p1}^1\Gamma_{23}^p-\Gamma_{p2}^2\Gamma_{23}^p-\Gamma_{p3}^3\Gamma_{23}^p\\
    &=-\frac{\partial \Gamma_{23}^1}{\partial x^1}-\frac{\partial \Gamma_{23}^2}{\partial x^2}+\Gamma_{03}^0\Gamma_{20}^0+\Gamma_{13}^0\Gamma_{20}^1+\Gamma_{23}^0\Gamma_{20}^2+\Gamma_{33}^0\Gamma_{20}^3\\
    &+\Gamma_{03}^1\Gamma_{21}^0+\Gamma_{13}^1\Gamma_{21}^1+\Gamma_{23}^1\Gamma_{21}^2+\Gamma_{33}^1\Gamma_{21}^3+\Gamma_{03}^2\Gamma_{22}^0+\Gamma_{13}^2\Gamma_{22}^1\\
    &+\Gamma_{23}^2\Gamma_{22}^2+\Gamma_{33}^2\Gamma_{22}^3+\Gamma_{03}^3\Gamma_{23}^0+\Gamma_{13}^3\Gamma_{23}^1+\Gamma_{23}^3\Gamma_{23}^2+\Gamma_{33}^3\Gamma_{23}^3\\
    &-\Gamma_{00}^0\Gamma_{23}^0-\Gamma_{10}^0\Gamma_{23}^1-\Gamma_{20}^0\Gamma_{23}^2-\Gamma_{30}^0\Gamma_{23}^3-\Gamma_{01}^1\Gamma_{23}^0-\Gamma_{11}^1\Gamma_{23}^1\\
    &-\Gamma_{21}^1\Gamma_{23}^2-\Gamma_{31}^1\Gamma_{23}^3-\Gamma_{02}^2\Gamma_{23}^0-\Gamma_{12}^2\Gamma_{23}^1-\Gamma_{22}^2\Gamma_{23}^2-\Gamma_{32}^2\Gamma_{23}^3\\
    &-\Gamma_{03}^3\Gamma_{23}^0-\Gamma_{13}^3\Gamma_{23}^1-\Gamma_{23}^3\Gamma_{23}^2-\Gamma_{33}^3\Gamma_{23}^3\\
    &=0
\end{align*}
\subsection{Ricci scalar}\label{subsecA2.2}
The Ricci scalar is the trace of the Ricci tensors that is obtained by using \eqref{eq:eleven}, \eqref{eq:seventeen}, \eqref{eq:eighteen}, \eqref{eq:nineteen} and \eqref{eq:twenty} as below:

\begin{align*}
    R=&g^{ik}R_{ik}=g^{00}R_{00}+g^{11}R_{11}+g^{22}R_{22}+g^{33}R_{33}\nonumber\\
    =&\left(e^{-2\nu}\right)\Big[-e^{2(\nu-\psi)}\left(\nu^{\prime\prime}+\nu^{\prime}\nu^{\prime}-\nu^{\prime}\psi^{\prime}+\mu_1^{\prime}\nu^{\prime}+\mu_2^{\prime}\nu^{\prime}\right)\nonumber\\
    &-e^{2(\nu-\mu_1)}\left(\Ddot{\nu}+\Dot{\nu}\Dot{\nu}-\Dot{\nu}\Dot{\mu_1}+\Dot{\psi}\Dot{\nu}+\Dot{\mu_2}\Dot{\nu}\right)\Big]\nonumber\\
    &-\left(e^{-2\psi}\right)\Big[\nu^{\prime\prime}+\mu_1^{\prime\prime}+\mu_2^{\prime\prime}+\left(\nu^{\prime}\right)^2+\left(\mu_1^{\prime}\right)^2+\left(\mu_2^{\prime}\right)^2-\nu^{\prime}\psi^{\prime}-\mu_1^{\prime}\psi^{\prime}-\mu_2^{\prime}\psi^{\prime}\nonumber\\
    &+e^{2(\psi-\mu_1)}\Big(\Ddot{\psi}+\Dot{\psi}\Dot{\psi}-\Dot{\psi}\Dot{\mu_1}+\Dot{\nu}\Dot{\psi}+\Dot{\mu_2}\Dot{\psi}\Big)\Big]\nonumber\\
    &-\left(e^{-2\mu_1}\right)\Big[e^{2(\mu_1-\psi)}\Big[\mu_1^{\prime\prime}-\mu_1^{\prime}\psi^{\prime}+\mu_1^{\prime}\mu_1^{\prime}+\mu_2^{\prime}\mu_1^{\prime}+\nu^{\prime}\mu_1^{\prime}\Big]\nonumber\nonumber\\
     &-\left(\Dot{\nu}\right)\left(\Dot{\mu_1}\right)+\Ddot{\nu}+\Ddot{\psi}+\Ddot{\mu_2}+\left(\Dot{\mu_2}\right)^2+\left(\Dot{\nu}\right)^2+\left(\Dot{\psi}\right)^2-\left(\Dot{\mu_2}\right)\left(\Dot{\mu_1}\right)-\left(\Dot{\psi}\right)\left(\Dot{\mu_1}\right)\Big]\nonumber\\
    &-\left(e^{-2\mu_2}\right)\Big[e^{2(\mu_2-\psi)}\Big[\mu_2^{\prime\prime}+\mu_2^{\prime}\mu_2^{\prime}-\mu_2^{\prime}\psi^{\prime}+\left(\nu^{\prime}\right)\left(\mu_2^{\prime}\right)+\left(\mu_1^{\prime}\right)\left(\mu_2^{\prime}\right)\Big]\nonumber\\
    &+e^{2(\mu_2-\mu_1)}\Big[\Ddot{\mu_2}+\Dot{\mu_2}\Dot{\mu_2}-\Dot{\mu_2}\Dot{\mu_1}+\left(\Dot{\nu}\right)\left(\Dot{\mu_2}\right)+\left(\Dot{\psi}\right)\left(\Dot{\mu_2}\right)\Big]\Big]
    \end{align*}
    \begin{align*}
    =&\Big[-e^{-2\psi}\left(\nu^{\prime\prime}+\nu^{\prime}\nu^{\prime}-\nu^{\prime}\psi^{\prime}+\mu_1^{\prime}\nu^{\prime}+\mu_2^{\prime}\nu^{\prime}\right)\nonumber\\
    &-e^{-2\mu_1}\left(\Ddot{\nu}+\Dot{\nu}\Dot{\nu}-\Dot{\nu}\Dot{\mu_1}+\Dot{\psi}\Dot{\nu}+\Dot{\mu_2}\Dot{\nu}\right)\Big]\nonumber\\
    &-\left(e^{-2\psi}\right)\Big[\nu^{\prime\prime}+\mu_1^{\prime\prime}+\mu_2^{\prime\prime}+\left(\nu^{\prime}\right)^2+\left(\mu_1^{\prime}\right)^2+\left(\mu_2^{\prime}\right)^2-\nu^{\prime}\psi^{\prime}-\mu_1^{\prime}\psi^{\prime}-\mu_2^{\prime}\psi^{\prime}\Big]\nonumber\\
    &-e^{-2\mu_1}\Big(\Ddot{\psi}+\Dot{\psi}\Dot{\psi}-\Dot{\psi}\Dot{\mu_1}+\Dot{\nu}\Dot{\psi}+\Dot{\mu_2}\Dot{\psi}\Big)\nonumber\\
    &-e^{-2\psi}\Big[\mu_1^{\prime\prime}-\mu_1^{\prime}\psi^{\prime}+\mu_1^{\prime}\mu_1^{\prime}+\mu_2^{\prime}\mu_1^{\prime}+\nu^{\prime}\mu_1^{\prime}\Big]\nonumber\\
     &-\left(e^{-2\mu_1}\right)\Big[-\left(\Dot{\nu}\right)\left(\Dot{\mu_1}\right)+\Ddot{\nu}+\Ddot{\psi}+\Ddot{\mu_2}+\left(\Dot{\mu_2}\right)^2+\left(\Dot{\nu}\right)^2+\left(\Dot{\psi}\right)^2
    \end{align*}

\begin{align}
&-\left(\Dot{\mu_2}\right)\left(\Dot{\mu_1}\right)-\left(\Dot{\psi}\right)\left(\Dot{\mu_1}\right)\Big]\nonumber\\
    &-e^{-2\psi}\Big[\mu_2^{\prime\prime}+\mu_2^{\prime}\mu_2^{\prime}-\mu_2^{\prime}\psi^{\prime}+\left(\nu^{\prime}\right)\left(\mu_2^{\prime}\right)+\left(\mu_1^{\prime}\right)\left(\mu_2^{\prime}\right)\Big]\nonumber\\
    &-e^{-2\mu_1}\Big[\Ddot{\mu_2}+\Dot{\mu_2}\Dot{\mu_2}-\Dot{\mu_2}\Dot{\mu_1}+\left(\Dot{\nu}\right)\left(\Dot{\mu_2}\right)+\left(\Dot{\psi}\right)\left(\Dot{\mu_2}\right)\Big]\nonumber\\
=&-e^{-2\psi}\Big[\nu^{\prime\prime}+\nu^{\prime}\nu^{\prime}-\nu^{\prime}\psi^{\prime}+\mu_1^{\prime}\nu^{\prime}+\mu_2^{\prime}\nu^{\prime}+\nu^{\prime\prime}+\mu_1^{\prime\prime}+\mu_2^{\prime\prime}+\left(\nu^{\prime}\right)^2\nonumber\\
   &+\left(\mu_1^{\prime}\right)^2+\left(\mu_2^{\prime}\right)^2-\nu^{\prime}\psi^{\prime}-\mu_1^{\prime}\psi^{\prime}-\mu_2^{\prime}\psi^{\prime}+\mu_1^{\prime\prime}-\mu_1^{\prime}\psi^{\prime}+\mu_1^{\prime}\mu_1^{\prime}\nonumber\\
   &+\mu_2^{\prime}\mu_1^{\prime}+\nu^{\prime}\mu_1^{\prime}+\mu_2^{\prime\prime}+\mu_2^{\prime}\mu_2^{\prime}-\mu_2^{\prime}\psi^{\prime}+\left(\nu^{\prime}\right)\left(\mu_2^{\prime}\right)+\left(\mu_1^{\prime}\right)\left(\mu_2^{\prime}\right)\Big]\nonumber\\
    &-e^{-2\mu_1}\Big[\Ddot{\nu}+\Dot{\nu}\Dot{\nu}-\Dot{\nu}\Dot{\mu_1}+\Dot{\psi}\Dot{\nu}+\Dot{\mu_2}\Dot{\nu}+\Ddot{\psi}+\Dot{\psi}\Dot{\psi}-\Dot{\psi}\Dot{\mu_1}+\Dot{\nu}\Dot{\psi}+\Dot{\mu_2}\Dot{\psi}\nonumber\\
     &-\left(\Dot{\nu}\right)\left(\Dot{\mu_1}\right)+\Ddot{\nu}+\Ddot{\psi}+\Ddot{\mu_2}+\left(\Dot{\mu_2}\right)^2+\left(\Dot{\nu}\right)^2+\left(\Dot{\psi}\right)^2-\left(\Dot{\mu_2}\right)\left(\Dot{\mu_1}\right)-\left(\Dot{\psi}\right)\left(\Dot{\mu_1}\right)\nonumber\\
    &+\Ddot{\mu_2}+\Dot{\mu_2}\Dot{\mu_2}-\Dot{\mu_2}\Dot{\mu_1}+\left(\Dot{\nu}\right)\left(\Dot{\mu_2}\right)+\left(\Dot{\psi}\right)\left(\Dot{\mu_2}\right)\Big]\nonumber\\
    =&-2e^{-2\psi}\Big[\nu^{\prime\prime}+\nu^{\prime}\nu^{\prime}-\nu^{\prime}\psi^{\prime}+\mu_1^{\prime}\nu^{\prime}+\mu_2^{\prime}\nu^{\prime}+\mu_1^{\prime\prime}+\mu_2^{\prime\prime}\nonumber\\
    &+\left(\mu_1^{\prime}\right)^2+\left(\mu_2^{\prime}\right)^2-\mu_1^{\prime}\psi^{\prime}-\mu_2^{\prime}\psi^{\prime}+\mu_2^{\prime}\mu_1^{\prime}\Big]\nonumber\\
    &-2e^{-2\mu_1}\Big[\Ddot{\nu}+\Dot{\nu}\Dot{\nu}-\Dot{\nu}\Dot{\mu_1}+\Dot{\psi}\Dot{\nu}+\Dot{\mu_2}\Dot{\nu}+\Ddot{\psi}+\Dot{\psi}\Dot{\psi}-\Dot{\psi}\Dot{\mu_1}+\Dot{\mu_2}\Dot{\psi}\nonumber\\
     &+\Ddot{\mu_2}+\left(\Dot{\mu_2}\right)^2-\left(\Dot{\mu_2}\right)\left(\Dot{\mu_1}\right)\Big]
\end{align}

\subsection{Einstein tensors}\label{subsecA2.3}
The $u$ and $\theta$ component of the Einstein tensors respectively are as follows:
\begin{align}
G_{11}=&R_{11}-\frac{1}{2}g_{11}R\nonumber\\
    =&\nu^{\prime\prime}+\mu_1^{\prime\prime}+\mu_2^{\prime\prime}+\left(\nu^{\prime}\right)^2+\left(\mu_1^{\prime}\right)^2+\left(\mu_2^{\prime}\right)^2-\nu^{\prime}\psi^{\prime}-\mu_1^{\prime}\psi^{\prime}-\mu_2^{\prime}\psi^{\prime}\nonumber\\
    &+e^{2(\psi-\mu_1)}\Big(\Ddot{\psi}+\Dot{\psi}\Dot{\psi}-\Dot{\psi}\Dot{\mu_1}+\Dot{\nu}\Dot{\psi}+\Dot{\mu_2}\Dot{\psi}\Big)\nonumber\\
    &-\frac{1}{2}\left(-e^{2\psi}\right)\Bigg[-2e^{-2\psi}\Big[\nu^{\prime\prime}+\nu^{\prime}\nu^{\prime}-\nu^{\prime}\psi^{\prime}+\mu_1^{\prime}\nu^{\prime}+\mu_2^{\prime}\nu^{\prime}+\mu_1^{\prime\prime}+\mu_2^{\prime\prime}\nonumber\\
    &+\left(\mu_1^{\prime}\right)^2+\left(\mu_2^{\prime}\right)^2-\mu_1^{\prime}\psi^{\prime}-\mu_2^{\prime}\psi^{\prime}+\mu_2^{\prime}\mu_1^{\prime}\Big]\nonumber\\
    &-2e^{-2\mu_1}\Big[\Ddot{\nu}+\Dot{\nu}\Dot{\nu}-\Dot{\nu}\Dot{\mu_1}+\Dot{\psi}\Dot{\nu}+\Dot{\mu_2}\Dot{\nu}+\Ddot{\psi}+\Dot{\psi}\Dot{\psi}-\Dot{\psi}\Dot{\mu_1}+\Dot{\mu_2}\Dot{\psi}\nonumber\\
     &+\Ddot{\mu_2}+\left(\Dot{\mu_2}\right)^2-\left(\Dot{\mu_2}\right)\left(\Dot{\mu_1}\right)\Big]\Bigg]\nonumber\\
     =&-\mu_1^{\prime}\nu^{\prime}-\mu_2^{\prime}\nu^{\prime}-\mu_2^{\prime}\mu_1^{\prime}-e^{2(\psi-\mu_1)}\Big[\Ddot{\nu}+\Dot{\nu}\Dot{\nu}-\Dot{\nu}\Dot{\mu_1}+\Dot{\mu_2}\Dot{\nu}\nonumber\\
     &+\Ddot{\mu_2}+\Dot{\mu_2}\Dot{\mu_2}-\Dot{\mu_2}\Dot{\mu_1}\Big]
\end{align}
and
\begin{align}
    G_{22}=&R_{22}-\frac{1}{2}g_{22}R\nonumber\\
    =&e^{2(\mu_1-\psi)}\Big[\mu_1^{\prime\prime}-\mu_1^{\prime}\psi^{\prime}+\mu_1^{\prime}\mu_1^{\prime}+\mu_2^{\prime}\mu_1^{\prime}+\nu^{\prime}\mu_1^{\prime}\Big]\nonumber\\
     &-\left(\Dot{\nu}\right)\left(\Dot{\mu_1}\right)+\Ddot{\nu}+\Ddot{\psi}+\Ddot{\mu_2}+\left(\Dot{\mu_2}\right)^2+\left(\Dot{\nu}\right)^2+\left(\Dot{\psi}\right)^2-\left(\Dot{\mu_2}\right)\left(\Dot{\mu_1}\right)-\left(\Dot{\psi}\right)\left(\Dot{\mu_1}\right)\nonumber\\
     &-\frac{1}{2}\left(-e^{2\mu_1}\right)\Bigg[-2e^{-2\psi}\Big[\nu^{\prime\prime}+\nu^{\prime}\nu^{\prime}-\nu^{\prime}\psi^{\prime}+\mu_1^{\prime}\nu^{\prime}+\mu_2^{\prime}\nu^{\prime}+\mu_1^{\prime\prime}+\mu_2^{\prime\prime}\nonumber\\
    &+\left(\mu_1^{\prime}\right)^2+\left(\mu_2^{\prime}\right)^2-\mu_1^{\prime}\psi^{\prime}-\mu_2^{\prime}\psi^{\prime}+\mu_2^{\prime}\mu_1^{\prime}\Big]\nonumber\\
    &-2e^{-2\mu_1}\Big[\Ddot{\nu}+\Dot{\nu}\Dot{\nu}-\Dot{\nu}\Dot{\mu_1}+\Dot{\psi}\Dot{\nu}+\Dot{\mu_2}\Dot{\nu}+\Ddot{\psi}+\Dot{\psi}\Dot{\psi}-\Dot{\psi}\Dot{\mu_1}+\Dot{\mu_2}\Dot{\psi}\nonumber\\
     &+\Ddot{\mu_2}+\left(\Dot{\mu_2}\right)^2-\left(\Dot{\mu_2}\right)\left(\Dot{\mu_1}\right)\Big]\Bigg]\nonumber\\
     =&-e^{2(\mu_1-\psi)}\Big[\nu^{\prime\prime}+\nu^{\prime}\nu^{\prime}-\nu^{\prime}\psi^{\prime}+\mu_2^{\prime}\nu^{\prime}+\mu_2^{\prime\prime}+\mu_2^{\prime}\mu_2^{\prime}-\mu_2^{\prime}\psi^{\prime}\Big]\nonumber\\
     &-\Big[\Dot{\psi}\Dot{\nu}+\Dot{\mu_2}\Dot{\nu}+\Dot{\mu_2}\Dot{\psi}\Big]
\end{align}

\section{Differential equation (48)}\label{secA3}
Substituting the expressions from equations \eqref{eq:six}, \eqref{eq:nine} and \eqref{eq:fortyseven} in the differential equation \eqref{eq:xfortyeight} we obtained:
\begin{align*}
    &\frac{\partial}{\partial u} \left[\Delta\frac{e^{\mu_2}}{e^{\nu}}\frac{\partial}{\partial u}\frac{e^{\nu}}{e^{\mu_2}}\right]+\frac{1}{\sin\theta}\frac{\partial}{\partial\theta} \left[\sin\theta\frac{e^{\mu_2}}{e^{\nu}}\frac{\partial}{\partial \theta}\frac{e^{\nu}}{e^{\mu_2}}\right]=0\nonumber\\
    \text{or\,\,\,\,} &\frac{\partial}{\partial u} \left[\Delta\frac{D(u^2+\eta^2)^{1/2}\sin\theta}{\frac{\sqrt{\Delta}}{D\left(u^2+\eta^2\right)^{1/2}}}\frac{\partial}{\partial u}\frac{\frac{\sqrt{\Delta}}{D\left(u^2+\eta^2\right)^{1/2}}}{D(u^2+\eta^2)^{1/2}\sin\theta}\right]\nonumber\\
    &\hspace{1cm}+\frac{1}{\sin\theta}\frac{\partial}{\partial\theta} \left[\sin\theta\frac{D(u^2+\eta^2)^{1/2}\sin\theta}{\frac{\sqrt{\Delta}}{D\left(u^2+\eta^2\right)^{1/2}}}\frac{\partial}{\partial \theta}\frac{\frac{\sqrt{\Delta}}{D\left(u^2+\eta^2\right)^{1/2}}}{D(u^2+\eta^2)^{1/2}\sin\theta}\right]=0\nonumber\\
    \text{or\,\,\,\,} &\frac{\partial}{\partial u} \left[\Delta^{1/2}D^2(u^2+\eta^2)\frac{\partial}{\partial u}\frac{\Delta^{1/2}}{D^2(u^2+\eta^2)}\right]+\frac{1}{\sin\theta}\frac{\partial}{\partial\theta} \left[\sin^2\theta D^2\frac{\partial}{\partial \theta}\frac{1}{D^2\sin\theta}\right]=0\nonumber\\
    \text{or\,\,\,\,} &\frac{\partial}{\partial u} \left[\Delta^{1/2}D^2(u^2+\eta^2)\left\{\frac{1}{D^2}\frac{\partial}{\partial u}\frac{\Delta^{1/2}}{(u^2+\eta^2)}+\frac{\Delta^{1/2}}{(u^2+\eta^2)}\frac{\partial}{\partial u}\frac{1}{D^2}\right\}\right]\nonumber\\
    &\hspace{1cm}+\frac{1}{\sin\theta}\frac{\partial}{\partial\theta} \left[\sin^2\theta D^2\left(\frac{1}{D^2}\frac{\partial}{\partial \theta}\frac{1}{\sin\theta}+\frac{1}{\sin\theta}\frac{\partial}{\partial \theta}\frac{1}{D^2}\right)\right]=0\nonumber\\
    \text{or\,\,\,\,} &\frac{\partial}{\partial u} \Bigg[\Delta^{1/2}D^2(u^2+\eta^2)\Bigg\{\frac{1}{D^2}\left(\frac{u-m}{(u^2+\eta^2)\Delta^{1/2}}-\frac{2u\Delta^{1/2}}{(u^2+\eta^2)^2}\right)\nonumber\\
    &+\frac{\Delta^{1/2}}{(u^2+\eta^2)}\frac{\partial}{\partial u}\frac{1}{D^2}\Bigg\}\Bigg]+\frac{1}{\sin\theta}\frac{\partial}{\partial\theta} \left[\sin^2\theta D^2\left(\frac{-1}{D^2}\frac{\cos\theta}{\sin^2\theta}+\frac{1}{\sin\theta}\frac{\partial}{\partial \theta}\frac{1}{D^2}\right)\right]=0
    \end{align*}
    \begin{align}
    \text{or\,\,\,\,} &\frac{\partial}{\partial u} \left[u-m-\frac{2u\Delta}{(u^2+\eta^2)}+\Delta D^2\frac{\partial}{\partial u}\frac{1}{D^2}\right]\nonumber\\
    &\hspace{3cm}+\frac{1}{\sin\theta}\frac{\partial}{\partial\theta} \left[-\cos\theta+\sin\theta D^2\frac{\partial}{\partial \theta}\frac{1}{D^2}\right]=0\nonumber\\
    \text{or\,\,\,\,} & 1-\frac{4u(u-m)}{(u^2+\eta^2)}+\frac{4u^2\Delta}{(u^2+\eta^2)^2}-\frac{2\Delta}{(u^2+\eta^2)}+\frac{\partial}{\partial u}\left(\Delta D^2\frac{\partial}{\partial u}\frac{1}{D^2}\right)\nonumber\\
    &\hspace{3cm}+ 1+\frac{1}{\sin\theta}\frac{\partial}{\partial\theta}\left(\sin\theta D^2\frac{\partial}{\partial \theta}\frac{1}{D^2}\right)=0\nonumber\\
    \text{or\,\,\,\,} & 1-\frac{4u(u-m)}{(u^2+\eta^2)}+\frac{4u^2\Delta}{(u^2+\eta^2)^2}-\frac{2\Delta}{(u^2+\eta^2)}+\Delta D^2\frac{\partial^2}{\partial u^2}\frac{1}{D^2}\nonumber\\
    &+\left(\frac{\partial}{\partial u}\frac{1}{D^2}\right)\frac{\partial}{\partial u}\left(\Delta D^2\right)+ 1+ D^2\frac{\partial^2}{\partial \theta^2}\frac{1}{D^2}+\frac{1}{\sin\theta}\left(\frac{\partial}{\partial\theta}\sin\theta D^2\right)\frac{\partial}{\partial \theta}\frac{1}{D^2}=0\nonumber\\
    \text{or\,\,\,\,} & \Delta D^2\frac{\partial^2}{\partial u^2}\frac{1}{D^2}+\Delta\left(\frac{\partial}{\partial u}\frac{1}{D^2}\right)\left(\frac{\partial}{\partial u} D^2\right)+2(u-m)D^2\frac{\partial}{\partial u}\frac{1}{D^2}\nonumber\\
     &+ D^2\frac{\partial^2}{\partial \theta^2}\frac{1}{D^2}+\left(\frac{\partial}{\partial\theta} D^2\right)\left(\frac{\partial}{\partial \theta}\frac{1}{D^2}\right)+D^2\cot\theta\frac{\partial}{\partial \theta}\frac{1}{D^2}\nonumber\\
     &+2-\frac{4u(u-m)}{(u^2+\eta^2)}+\frac{4u^2\Delta}{(u^2+\eta^2)^2}-\frac{2\Delta}{(u^2+\eta^2)}=0\end{align}




\end{appendices}

\end{document}